\documentclass[aps,prb,twocolumn,superscriptaddress]{revtex4-1}
\usepackage{graphicx}
\usepackage[hidelinks]{hyperref} 
\usepackage{bm}
\usepackage{amsmath}
\usepackage{lipsum}

\usepackage{hyperref}
\hypersetup{colorlinks,
}

\graphicspath{{./}}

\begin{document}


\title{Crystal-electric-field excitations in a quantum-spin-liquid candidate NaErS$_2$ }

\renewcommand*{\thefootnote}{\arabic{footnote}}

\author{Shang Gao}
\email[]{shang.gao@riken.jp}
\thanks{Current address: Materials Science \& Technology Division and Neutron Science Division, Oak Ridge National Laboratory, Oak Ridge, TN 37831, USA}
\affiliation{RIKEN Center for Emergent Matter Science, Wako 351-0198, Japan}

\author{Fan Xiao}
\email[]{fan.xiao@psi.ch}
\affiliation{Laboratory for Neutron Scattering and Imaging, Paul Scherrer Institut, CH-5232 Villigen PSI, Switzerland}
\affiliation{Department of Chemistry and Biochemistry, University of Bern, Freiestrasse 3, Bern, Switzerland}

\author{Kazuya Kamazawa}
\affiliation{Neutron Science and Technology Center, Comprehensive Research Organization for Science and Society, Tokai, Ibaraki, 319-1106, Japan}

\author{Kazuhiko Ikeuchi}
\affiliation{Neutron Science and Technology Center, Comprehensive Research Organization for Science and Society, Tokai, Ibaraki, 319-1106, Japan}

\author{Daniel Biner}
\affiliation{Department of Chemistry and Biochemistry, University of Bern, Freiestrasse 3, Bern, Switzerland}

\author{Karl W. Kr\"amer}
\affiliation{Department of Chemistry and Biochemistry, University of Bern, Freiestrasse 3, Bern, Switzerland}

\author{Christian R\"uegg}
\affiliation{Research Division Neutrons and Muons, Paul Scherrer Institut, CH-5232 Villigen PSI, Switzerland}
\affiliation{Department of Quantum Matter Physics, University of Geneva, CH-1211 Geneva, Switzerland}

\author{Taka-hisa Arima}
\affiliation{RIKEN Center for Emergent Matter Science, Wako 351-0198, Japan}
\affiliation{Department of Advanced Materials Science, University of Tokyo, Kashiwa 277-8561, Japan}

\date{\today}


\begin{abstract}

The delafossite family of compounds with a triangular lattice of rare earth ions has been recently proposed as a candidate host for quantum spin liquid (QSL) states. To realize QSLs, the crystal-electric-field (CEF) ground state of the rare earth ions should be composed of a doublet that allows sizable quantum tunneling, but till now the knowledge on CEF states in the delafossite compounds is still limited. Here we employ inelastic neutron scattering (INS) to study the CEF transitions in a powder sample of the delafossite NaErS$_2$, where the large total angular momentum $J = 15/2$ of the Er$^{3+} $ ions and the resulting plethora of CEF transitions enable an accurate fit of the CEF parameters. Our study reveals nearly isotropic spins with large $J_z = \pm 1/2$ components for the Er$^{3+}$ CEF ground states, which might facilitate the development of a QSL state. The scaling of the obtained CEF Hamiltonian to different rare earth ions suggests that sizable $J_z = \pm 1/2$ components are generally present in the CEF ground states, supporting the ternary sulfide delafossites as potential QSL hosts.

\end{abstract}

\maketitle

\section{Introduction}
The QSL state, where the conventional magnetic long-range order (LRO) is completely removed by quantum fluctuations, has been fascinating physicists since it was proposed in the 1970s~\cite{fazekas_1974_on}. Similar to the well-known cases of one-dimensional spin chains~\cite{giamarchi_quantum_2003}, the fundamental excitations in QSLs are fractional spin-1/2 excitations called spinons, which can be either gapped or gapless depending on the specific system~\cite{normand_frontiers_2009, balents_spin_2010, savary_quantum_2016, zhou_quantum_2017}. Theoretical investigations have revealed the spinons in some QSLs to be highly entangled with each other, leading to fractional statistics and exotic braiding properties that might be utilized for topological quantum computing~\cite{kitaev_anyons_2006,nayak_non_2008}.

The initial quest for QSLs was focused on intrinsic spin-1/2 systems such as the Cu$^{2+}$-based compounds~\cite{zhou_quantum_2017, shimizu_spin_2003}. One prominent example is the herbertsmithite ZnCu$_3$(OH)$_6$Cl$_2$~\cite{mendels_quantum_2007, han_fractionalized_2012, fu_evidence_2015}. In this compound, the Cu$^{2+}$ ions form a two-dimensional (2D) kagom\'e lattice with geometric frustration. Using INS, an almost ‘featureless’ excitation continuum was revealed~\cite{han_fractionalized_2012}, which is consistent with the spinon excitations and, more importantly, demonstrates that QSL can exist in real materials.

Recently, the search for QSL candidates has been extended to the rare earth systems. This is surprising at the first glance, because the rare earth ions usually have a relatively large angular momentum $J$, which disfavors quantum fluctuations. However, with an appropriate CEF,  the ground state doublet of the rare earth ions might have considerable components of $|J, J_z\rangle$ with a relatively small $|J_z|$ that allows quantum tunnelling~\cite{rau_magnitude_2015, rau_frustration_2018}. If this ground state is well separated from the excited states, the spin degree-of-freedom of the rare-earth ions will effectively behave as spin-1/2. One of the best-known examples is the quantum spin ice state realized in the rare-earth pyrochlores~\cite{hermele_pyrochlore_2004,gingras_quantum_2014}. For the Dy and Ho-based pyrochlore systems with only relatively large $|J_z|$ components in the CEF ground state, a classical spin ice state is realized, where each tetrahedron has a two-in-two-out spin configuration~\cite{bramwell_spin_2001, fennell_magnetic_2009}. While for the Tb, Yb, and Pr-based pyrochlores~\cite{fennell_power_2012, chang_higgs_2012, sibille_experimental_2018}, a relatively high magnitude of quantum spin tunnellings is observed, which drives the classical spin ice state into a QSL state with emergent U(1) quantum electrodynamics~\cite{hermele_pyrochlore_2004, benton_seeing_2012}.

Given the success of the effective spin-1/2 picture in the rare-earth pyrochlores, it is natural and tempting to advance this concept to other frustrated lattices, especially the 2D triangular lattice where the idea of QSL was originally conceived~\cite{fazekas_1974_on}. According to theoretical calculations~\cite{li_anisotropic_2016}, the effective spin-1/2 Hamiltonian for rare earth spins on a triangular lattice might contain transverse coupling terms that can induce competing ground states in the classical solution, whereupon a QSL state could emerge near the phase boundary once quantum fluctuations are included. Following this argument, the triangular lattice compound YbMgGaO$_4$ has recently been proposed as a candidate host for the QSL state~\cite{li_rare_2015, paddison_continuous_2016, shen_evidence_2016}. However, due to the Mg-Ga disorder that is intrinsic in this compound, it is unclear whether the broad excitations that have been observed in INS are due to quantum fluctuations or disorder effects~\cite{xu_absence_2016, zhu_disorder_2017, kimchi_valence_2018}.

\begin{figure}[t!]
\includegraphics[width=0.45\textwidth]{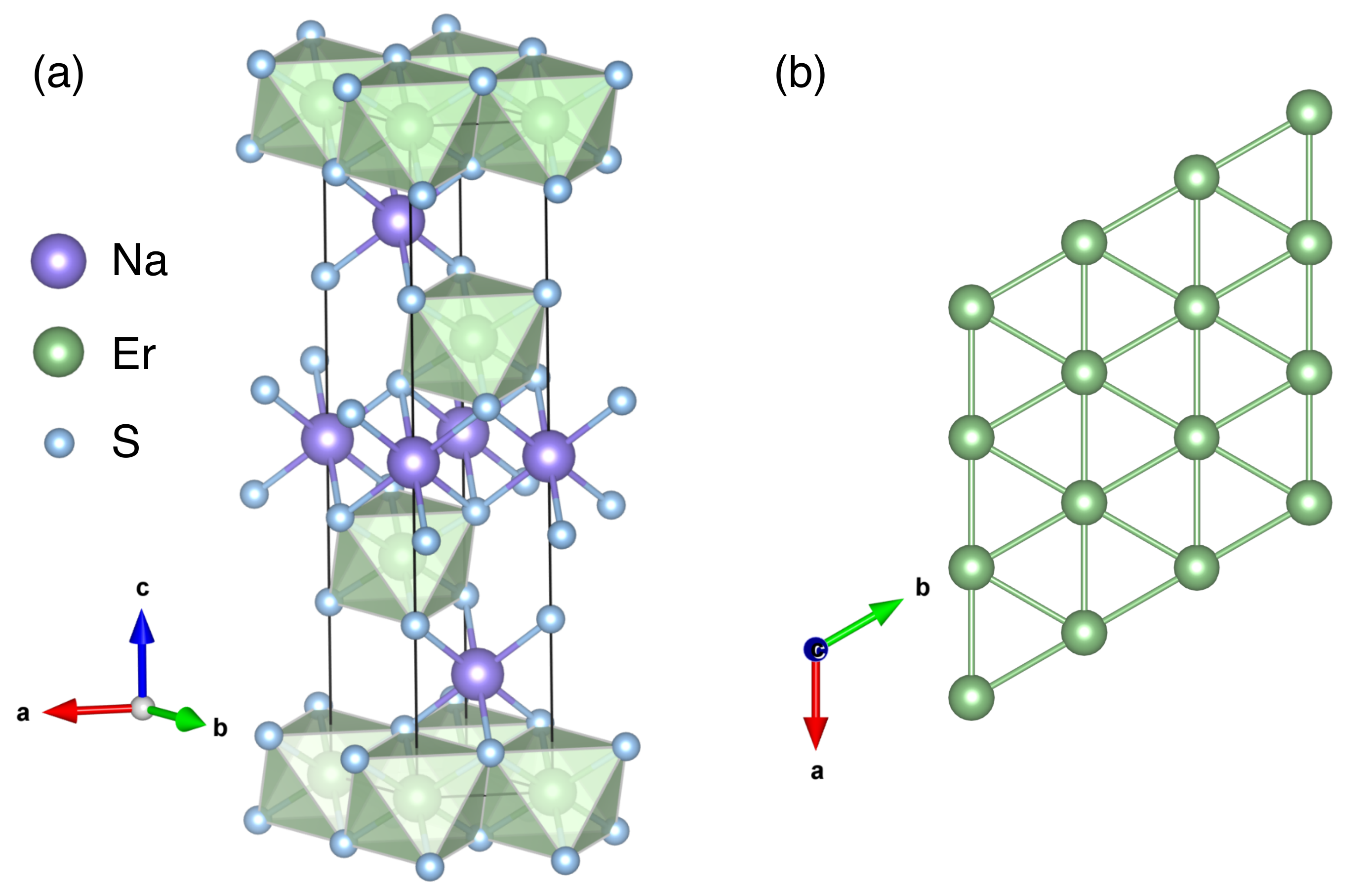}
\caption{(a) Crystal structure of NaErS$_2$~\cite{schleid_single_1993}. The Na$^{+}$ and Er$^{3+}$ ions occupy the $3b$ and $3a$ sites, respectively. The Er-S octahedra are explicitly shown. (b) The triangular lattice formed by the Er$^{3+}$ ions viewed along the $c$ axis.
\label{fig:str}}
\end{figure}

The delafossite family of compounds $A$Ln$X_2$, where Ln are rare earth ions, $A$ = Na, K, Cu(I), and $X$ = O, S, Se, might be the sought-after QSL candidates that are free from any disorder~\cite{liu_rare_2018, hashimoto_magnetic_2003, dong_structure_2008, miyasaka_synthesis_2009}. Similar to the parent delafossite mineral CuFeO$_2$, $A$Ln$X_2$ crystallize in the space group $R$\={3}$m$~\cite{schleid_single_1993}, with both $A$ and Ln sites forming triangular lattices as shown in Fig.~\ref{fig:str}. Especially, the Ln$X_2$ layers consist of Ln ions with $D_{3d}$ site symmetry located at the center of edge-sharing $X$-octahedra, similar to YbMgGaO$_4$. Detailed experimental studies on the magnetic properties of rare earth delafossites have been reported for the Yb-based compounds, including NaYbS$_2$~\cite{baenitz_planar_2018, sichelschmidt_electron_2019} and NaYbO$_2$~\cite{bordelon_field_2019, ranjith_field_2019, ding_gapless_2019}, which revealed the absence of magnetic LRO in both compounds and suggested possible QSL states. 

In order to facilitate the QSL search in the delafossites, it is crucial to have an overview of their CEF environment. As exemplified by the spin ice compounds~\cite{gardner_magnetic_2010, bertin_crystal_2012, gao_dipolar_2018, reig_neutron_2019}, the CEF parameters in systems with similar structures normally obey the scaling rule. Therefore, compared to the Yb$^{3+}$ ions with $J = 7/2$~\cite{baenitz_planar_2018, ding_gapless_2019}, rare earth ions with a larger $J$ allow more CEF transitions, which will enable a more accurate fit of the CEF parameters and thus provide a reference in the study of the similar delafossite compounds.

Here we report INS investigations on the CEF transitions in NaErS$_2$, where the Er$^{3+}$ has a total angular momentum of $J = 15/2$. Our studies reveal nearly isotropic spins with large $J_z = \pm 1/2$ components for the Er$^{3+}$ CEF ground state doublet that allow spin quantum tunnelling~\cite{rau_magnitude_2015, rau_frustration_2018}. The scaling of the obtained CEF Hamiltonian to different rare earth ions will foster the search for QSL states in the sulfide delafossites.

\begin{figure}[t!]
\includegraphics[width=0.48\textwidth]{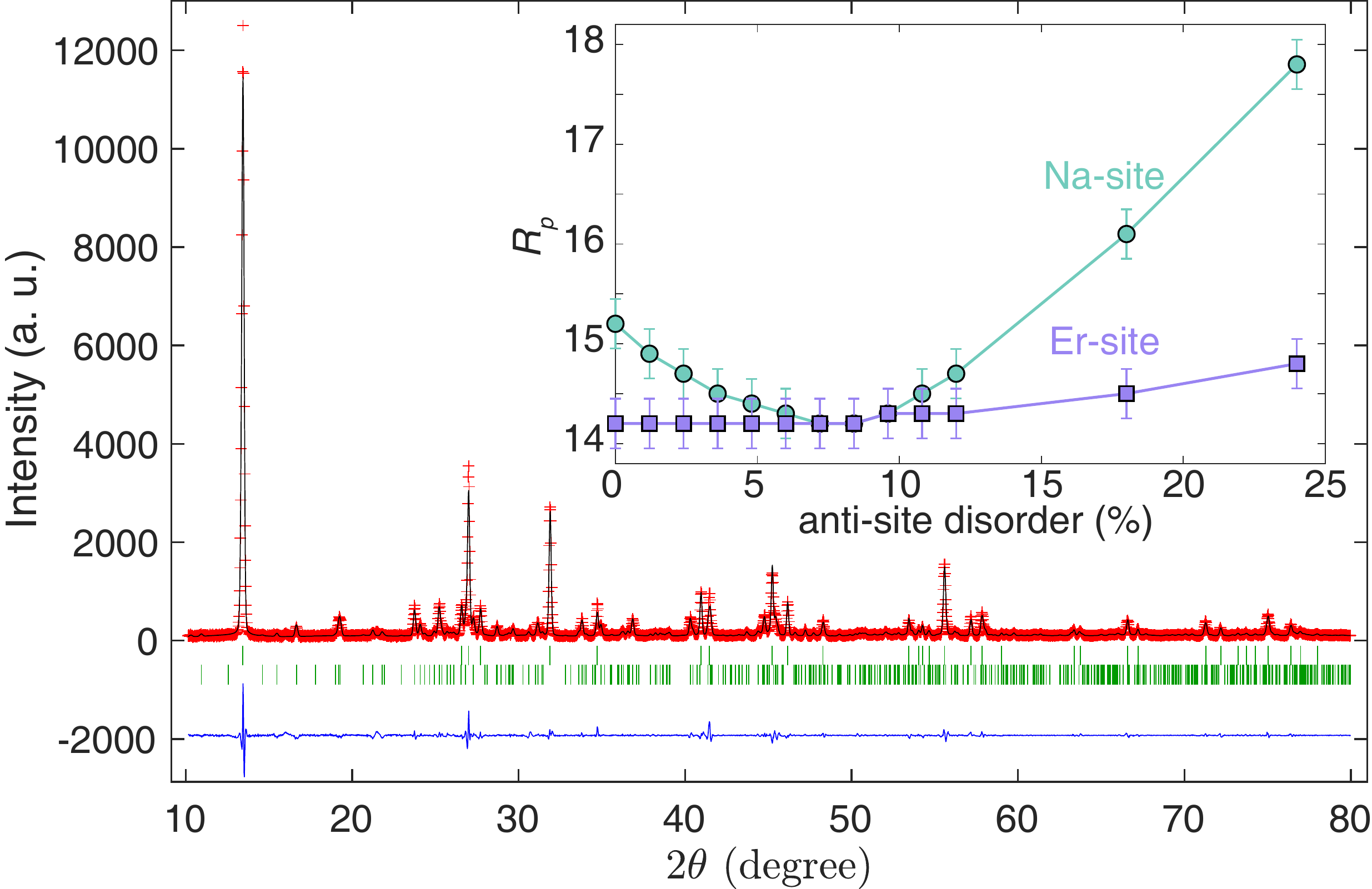}
\caption{Refinement results of the x-ray diffraction data measured at room temperature for NaErS$_2$ polycrystalline sample. Data points are shown as red crosses. The calculated pattern is shown as the black solid line. The upper and lower vertical bars show the positions of the Bragg peaks for NaErS$_2$ and Er$_2$S$_3$, respectively. The blue line at the bottom shows the difference of measured and calculated intensities. Inset shows the $R_p$ factor as a function of the anti-site disorder at the Na and Er sites.
\label{fig:xrd}}
\end{figure}

\begin{figure*} [t]
\includegraphics[width=0.75\textwidth]{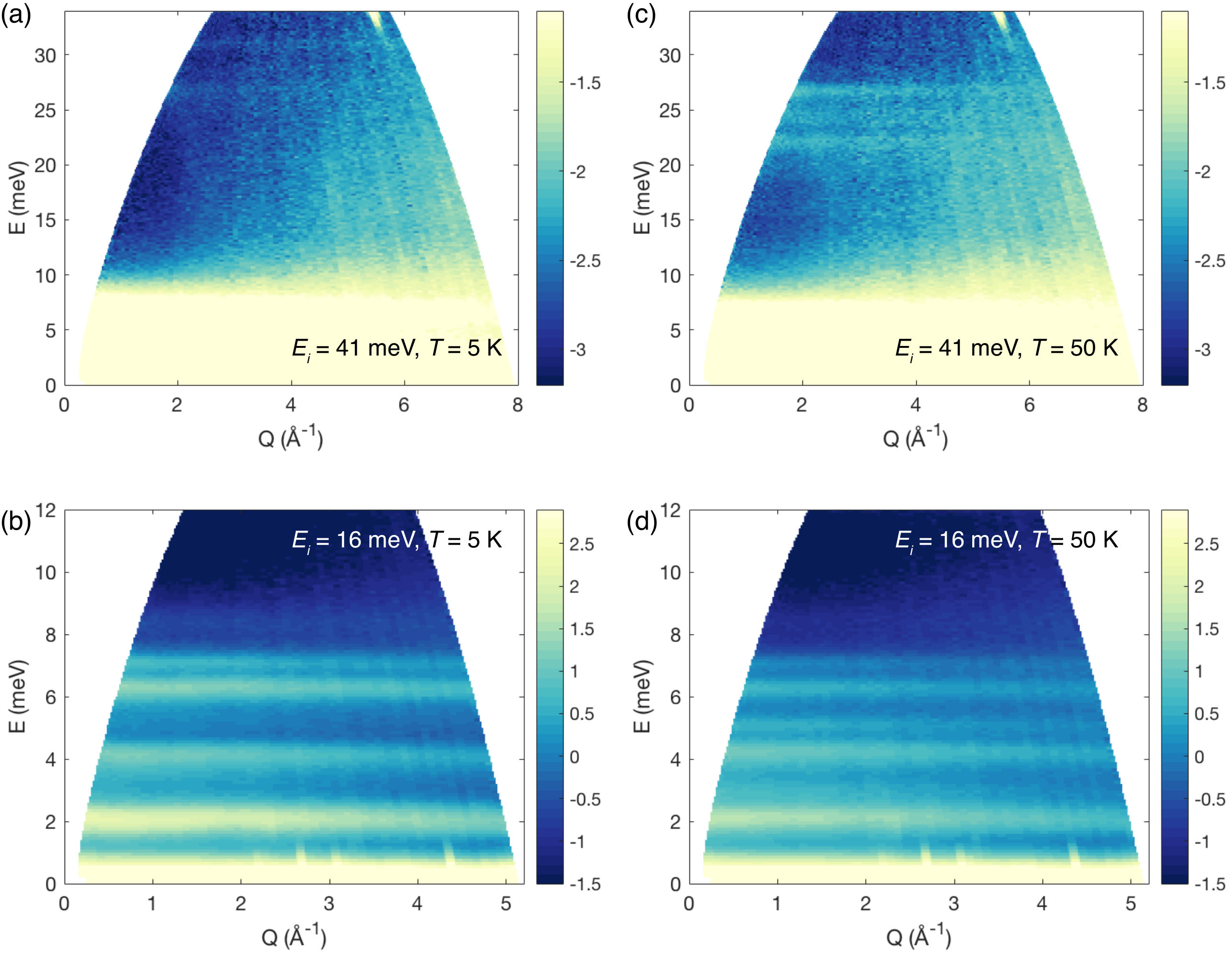}
\caption{INS spectra $S(Q,\omega)$ of a NaErS$_2$ powder sample collected on 4SEASONS at temperatures $T=5$~K (panels a and b) and 50 K (panels c and d), with incoming neutron energy $E_i=16$ meV (panels b and d) and 41 meV (panels a and c). At elevated temperatures, additional excitations originating from thermally populated doublets are observed.
\label{fig:map}}
\end{figure*}

\section{Experimental details}\label{sec:exp_det}
Polycrystalline samples of NaErS$_2$ were prepared using the solid state method proposed by Schleid \emph{et al} ~\cite{schleid_single_1993}, in which NaCl served as both reagent and flux. Under N$_2$, Er grains, sulfur, and NaCl in a molar ratio of 2:3:9 were loaded into a Ta ampoule, which was sealed by arc-welding under He. The Ta ampoule was sealed in a silica ampoule under vacuum, slowly heated up to 850$^\circ$C with 20$^\circ$C/h and kept for 7 days before cooling down to room temperature. The final product was rinsed with H$_2$O and acetone several times to remove water-soluble Na$_3$ErCl$_6$. 

Powder x-ray diffraction (XRD) patterns were measured on a STOE STADIP diffractometer in reflection (Bragg-Brentano) geometry in air at room temperature. Diffraction patterns with Cu $K_{\alpha 1}$ radiation ($\lambda$ = 1.54059~\AA) from a focusing $\alpha$-SiO$_2$ (101) monochromator were recorded on a linear position-sensitive detector with 0.01$^\circ$ resolution in 2$\theta$. Rietveld refinement was performed in the $R$\={3}$m$ space group using the FULLPROF program~\cite{rodriguez_recent_1993}.

INS experiments were performed on the 4SEASONS time-of-flight (TOF) spectrometer at the Materials and Life Science Experimental Facility MLF of J-PARC in Japan~\cite{kajimoto_fermi_2011}. The setup with a radial collimator and a neutron beam size of $20\times20$ mm$^2$ was employed. A NaErS$_2$ powder sample of 1.8~g was packed in an envelope of aluminum foil, curled up and installed in an aluminium sample can with outer/inner diameter of 20.5/20.0 mm. This configuration reduced the neutron absorption caused by the Er isotopes in the sample~\cite{ruminy_crystal_2016, gao_dipolar_2018}. For our measurements, the chopper frequency was set to 300 Hz, and the repetition rate multiplication method~\cite{nakamura_first_2009} allows the measurement with multiple incident energies of $E_i$ = 222, 80, 41, 24.7, 16, and 12 meV to be collected at the same time. A GM refrigerator was mounted to reach temperatures between 5 and 250 K. Besides the NaErS$_2$ sample, measurements were also performed on a vanadium standard to allow a quantitative comparison for data collected at 300 K with the same instrumental setup. The acquired data were analyzed with the UTSUSEMI software package~\cite{inamura_development_2013}.

\section{Results}\label{sec:results}

The XRD pattern for our NaErS$_2$ sample is shown in Fig.~\ref{fig:xrd}. The refined lattice parameters $a = 3.93343(4)$~\AA\ and $c =19.8378(2)$ \AA\ are in good agreement with the published crystal structure~\cite{schleid_single_1993}. NaErS$_2$ crystallizes in space group $R\overline{3}m$ with Na$^+$ and Er$^{3+}$ ions on sites $3b$ (0 0 0.5) and $3a$ (0 0 0), repectively. The S$^{2-}$ ions occupy the site $6c$ (0 0 $z$) with $z=0.2461(4)$. A satisfactory fit was obtained by including a preferred orientation along the [001] direction due to the plate-like habit of the NaErS$_2$ polycrystals. The $R$-factors are $R_p = 14.2\ \%$, $R_{wp}  = 15.8\ \%$, and $\chi^2 = 1.9$. The inset of Fig.~\ref{fig:xrd} presents the value of the $R_p$ factor as a function of the anti-site disorder at the Na and Er sites. Although the Er sites are fully occupied within our experimental resolution, a small fraction of 7~\% anti-site disorder is discerned at the Na sites. This disorder on the Na sites might cause the tail-like broadening in the CEF excitations as discussed in the following section.

A secondary phase is observed in all the synthesized batches, which can be assigned to Er$_{2+x}$S$_{3+y}$ impurities and has been treated with the Le Bail profile fit assuming a $P2_1/m$ space group. Using the strongest reflections for NaErS$_2$ at $2\theta \sim 13^{\circ}$ and for the secondary phase at $2\theta \sim 25^{\circ}$, the fraction of the secondary phase is estimated to be $\sim 5\%$. 

Fig.~\ref{fig:map} summarizes the NaErS$_2$ neutron spectra collected at $T = 5$ and 50~K with $E_i = 16$ and 41~meV. The strong intensity spot in the $E_i = 41$~meV spectra at wavevector transfer $Q = 5.5 $~\AA$^{-1}$ and energy transfer $E=34$ meV is spurious due to unshielded scattered neutrons from the beam catcher. For TOF neutron spectrometers, the energy resolution scales with the incoming neutron energy and can be estimated by the full-width-half-maximum (FWHM) of the incoherent scattering in the vanadium standard measurements. In our experiment, the energy resolution was estimated to be 0.50, 0.80, and 2.51 meV for $E_i = $ 12, 16, and 41~meV, respectively. Therefore, a relatively high $E_i$ of~41 meV allow access to the high energy excitations, while a relatively low $E_i$ of 16 or 12 meV resolves the different excitations at low energies.

At $T = 5$ K, four dispersionless excitations are observed at around 2.0, 4.0, 6.0, and 6.8 meV in the $E_i = 16$ meV spectra shown in Fig.~\ref{fig:map}(b), and three  relatively weak excitations can be discerned at 26.5, 28.3, 30.9 meV in the $E_i = 41$ meV spectra shown in Fig.~\ref{fig:map}(a). In $D_{3d}$ symmetry, the Er$^{3+}$ $^4I_{15/2}$ manifold splits into eight Kramers doublets. Therefore, the seven excitations observed in our INS spectra can be ascribed to the Stokes transitions from the CEF ground state doublet to the seven excited doublets. At an elevated temperature of 50~K, the excited doublets are thermally populated, leading to two additional transitions at $\sim 2.9$ and 5.0~meV in Fig.~\ref{fig:map}(d) and three high-energy transitions at 22.0, 24.0, and 26.5 meV in Fig.~\ref{fig:map}(c). Transitions at $\sim 2.9$, 22.0, 24.0, and 26.5 meV are due to the excitations from the doublet at $\sim 2.0$ meV, while the transition at $\sim 5.0$ meV is due to the excitations from the doublet at $\sim 4.0$ meV.

\begin{figure}[t!]
\includegraphics[width=0.48\textwidth]{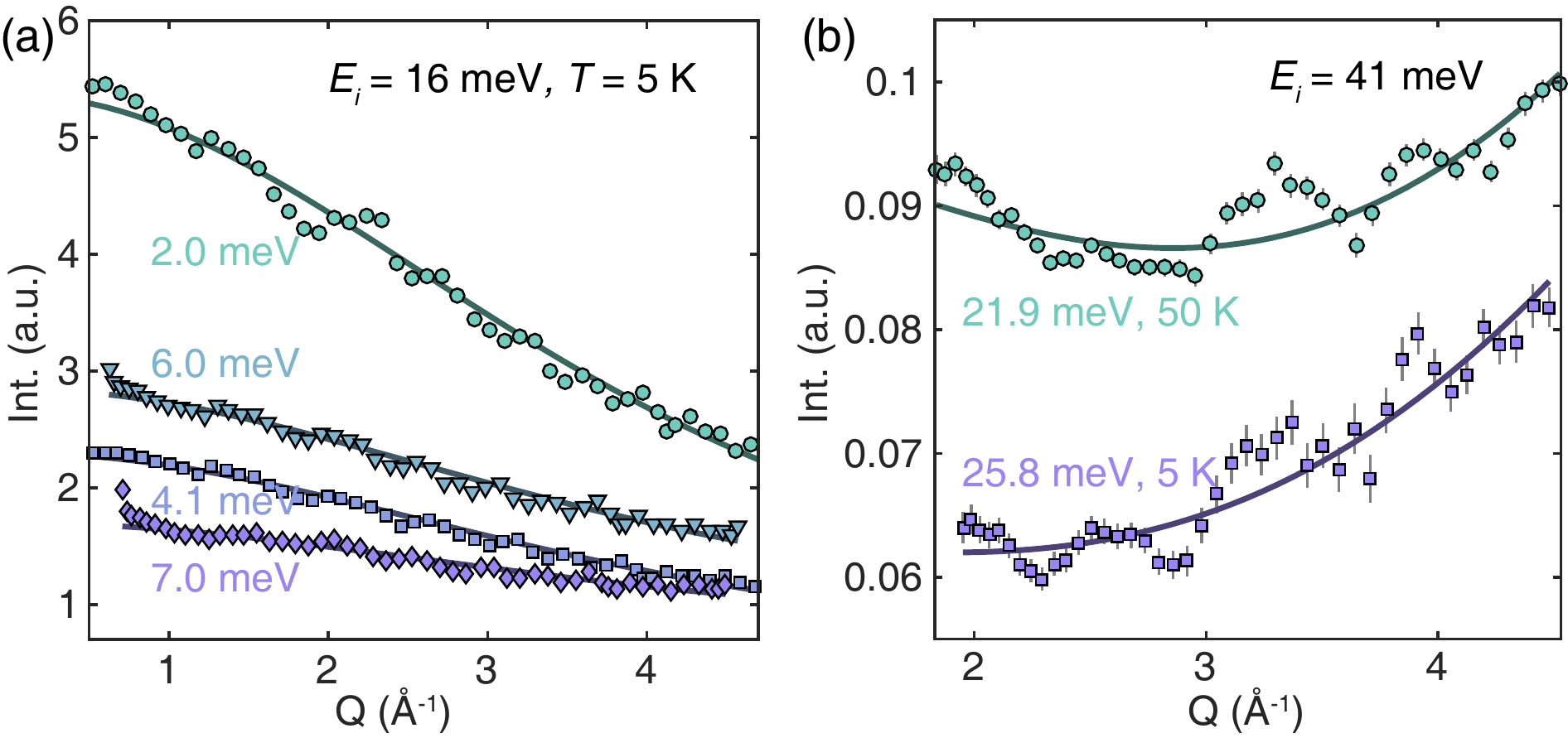}
\caption{(a) Momentum transfer dependence of the CEF excitations with $E_i = 16$~meV at $T=5$ K. Data points represent intensities integrated within an energy range of 1.2 meV centered around 2.0 meV (green circles), 4.1 meV (purple squares), 6.0 meV (green triangles), and 7.0 meV (purple diamonds). Solid lines are the corresponding fits by the magnetic form factor of the Er$^{3+}$ ions plus a flat background to confirm the magnetic origin of the excitations. (b) Comparison for the momentum transfer dependence of the CEF excitations with $E_i = 41$~meV. Purple squares (green cricles) are intensities integrated within an energy range of 2.5 (1.6) meV centered around 25.8 (21.9) meV measured at 5 (50)~K. Solid lines are the corresponding fits by the magnetic form factor of the Er$^{3+}$ ions plus a flat background together with a $Q^2$ term.
\label{fig:qdep}}
\end{figure}

INS probes the CEF transitions though the dipolar interactions between the neutron and electron spins (see Eqn.~(\ref{eq:ins})). Therefore, the neutron scattering length for CEF transitions should be proportional to the magnetic form factor $f(Q)$ of the Er$^{3+}$ ions that is monotonously decreasing with $Q$. Fig.~\ref{fig:qdep}(a) plots the $Q$-dependence of the INS intensities integrated at around $E = 2.0$, 4.1, 6.0, and 7.0 meV within an energy width of 1.2~meV. The integrated intensities decrease monotonously with $Q$, and can be fitted by the square of the form factor $f^2(Q)$ plus a constant background, which confirms the CEF origin of these excitations. In contrast, the high-$E$ modes observed at $T$ = 5 K shown in Fig.~\ref{fig:qdep}(b) exhibit a $Q$-quadratic behavior that is typical for phonon excitations. At 50 K, the CEF contributions to the 4.0~meV $\rightarrow$~25.8~meV transition become more obvious, leading to a non-monotonous $Q$-dependence for the 21.9-meV mode that can be described by the Er$^{3+}$ form factor plus a $Q^2$ term. Therefore, the high-$E$ modes should have contributions from both the CEF and phonon excitations.

The energies of the CEF levels together with their INS intensities can be quantitatively analyzed using the CEF Hamiltonian. As noted by Hutchings in the 1960s~\cite{hutchings_point_1964}, different normalization schemes exist for the CEF operators,  leading to different conventions in the CEF Hamiltonian definition. In the Stevens convention, the Hamiltonian is usually written as $\mathcal{H} = \sum_{l,m} B_l^m \hat{O}_l^m$, where the normalization factors for the CEF operators $\hat{O_l^m}$, or the so-called Stevens factors, are implicitly included in the CEF parameters $B_l^m$. Here the integer $l$ ranges from 0 to 6 for $f$-electrons, and the integer $m$ ranges from $-l$ to $l$. In the Wybourne convention, the Hamiltonian can be written as $\mathcal{H} = \sum_{l,m} L_m^l \hat{C}_m^l$, with the Stevens factors included in the CEF operators $\hat{C}_m^l$ instead of the  CEF parameters $L_m^l$. Here we follow the Wybourne convention and introduce the CEF operators $\hat{T}_l^m = \hat{C}_{-m}^l + (-1)^m\hat{C}_m^l$ for $m \geq 0$ as implemented in the McPhase program~\cite{rotter_using_2004}. For rare-earth ions with $D_{3d}$ symmetry, the CEF Hamiltonian becomes:
\begin{equation}
\label{eq:cef}
\mathcal{H} = L_2^0\hat{T}_2^0 + L_4^0\hat{T}_4^0 + L_4^3\hat{T}_4^3
					 + L_6^0\hat{T}_6^0 + L_6^3\hat{T}_6^3 + L_6^6\hat{T}_6^6\ \textrm{,}
\end{equation}
where the $z$ direction is along the three-fold rotation axis. The CEF parameters thus defined are related to the original Wybourne CEF parameters by a factor of $(-1)^m$.

Due to the large separation of $\sim800$~meV between the low-energy manifold $^4I_{15/2}$ and the higher-energy manifolds for isolated Er$^{3+}$ ions, we diagonalize the CEF Hamiltonian in the Hilbert space spanned by the basis vectors $|J=15/2, J_z\rangle$ within the $^4I_{15/2}$ manifold. The INS cross section for the CEF excitations on powder sample is then expressed as~\cite{rosenkranz_crystal_2000, xu_magnetic_2015}
\begin{equation}
\label{eq:ins}
\frac{d^{2} \sigma}{d \Omega d E}=c f^{2}(Q) \frac{k_f}{k_i} \sum_{\alpha} \sum_{i,f} p_i |\langle f|\hat{J}_{\alpha}| i\rangle|^{2}\delta(E_i-E_f+E) \ \textrm{,}
\end{equation}
where $c$ is a constant, $|i\rangle$ and $|f\rangle$ are the eigenfunctions of the CEF Hamiltonian and represent the initial and final wavefunctions, repectively. $E_i$ ($k_i$) and $E_f$ ($k_f$) are the energies (wavevectors) of the incoming and scattered neutrons, respectively. The occupation probability $p_i$ for the state at $E_i$ is described by the Boltzmann distribution $p_i = \exp{(-E_i/kT)}/\sum_i \exp{(-E_i/kT)}$. $\hat{J}_\alpha$ with $\alpha = x$, $y$, and $z$ are the angular momentum operators. $\delta(E_i-E_f+E)$ is the delta function.

\begin{table}[b!]
\caption{Fitted Wybourne CEF parameters (meV) for Er$^{3+}$ in NaErS$_2$. Errors are conservative estimates based on repeated Monte Carlo simulations.}
\label{tab:cef}
\centering
\begin{tabular}{cccccc}
\toprule
$L_2^0$ & $L_4^0$ & $L_4^3$ & $L_6^0$ & $L_6^3$ & $L_6^6$ \\
 \hline
$-24.7(1)$ & $-76.8(4)$ & $-128.0(7)\, $ & $29.3(1)$ & $-0.1(1)\,$ & $24.6(1)$\\
\botrule
\end{tabular}
\end{table}

\begin{figure}[t!]
\includegraphics[width=0.48\textwidth]{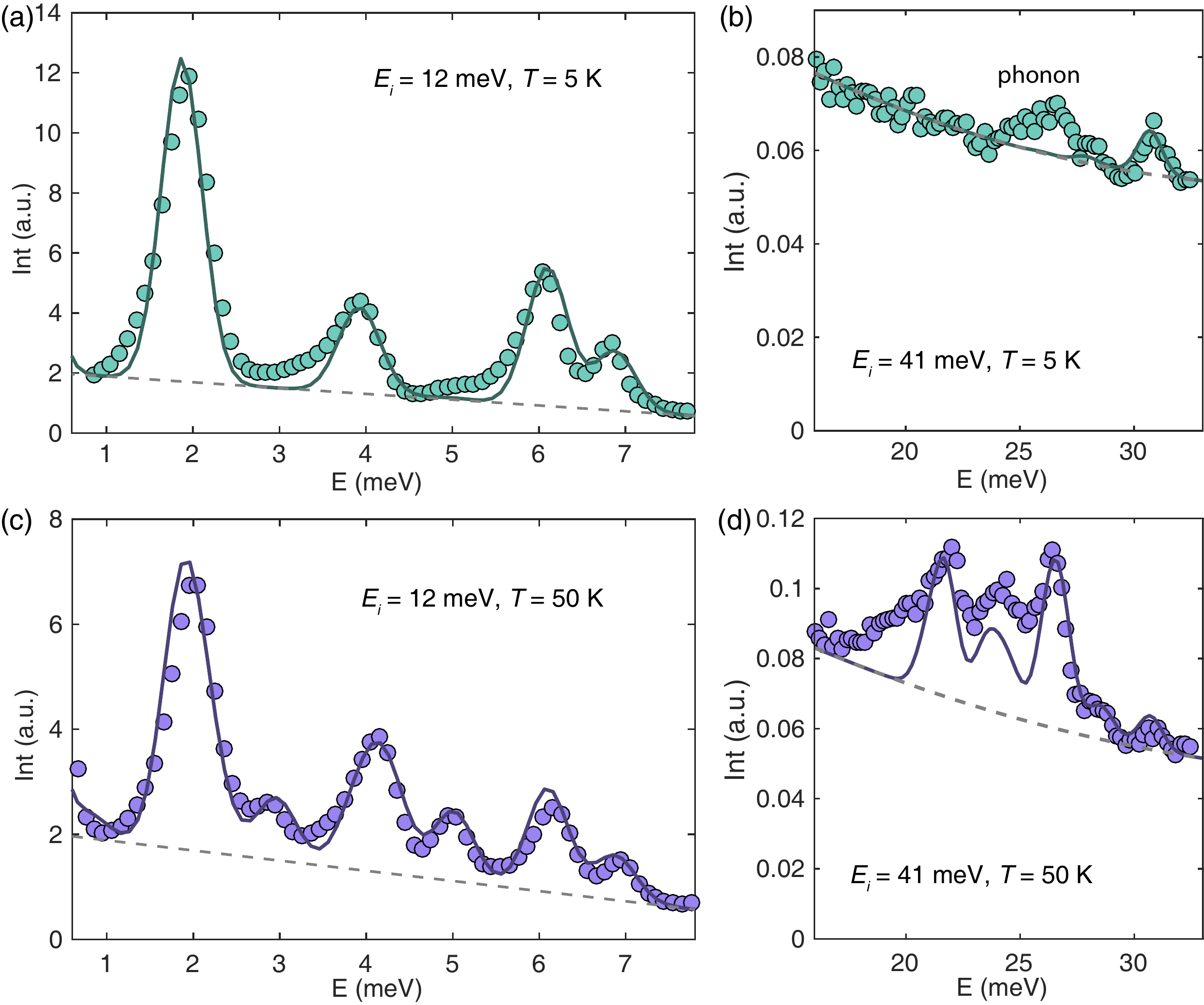}
\caption{CEF excitations collected at $T = 5$~K (panels a and b) and 50~K (panels c and d) with incoming neutron energy of $E_i = 12$~meV (panels a and c) and 41~meV (panels b and d). In panels a and c (b and d), data points represent intensities integrated within a momentum transfer range of $1.2\sim2.2$ ($2.2\sim3.2$)~\AA$^{-1}$. Solid lines are the corresponding fits using the CEF Hamiltonian plus a polynomial background term shown as the dashed lines. Errorbars representing standard deviations are smaller than the symbol size. The fitted CEF parameters are shown in Table~\ref{tab:cef}.
\label{fig:fit}}
\end{figure}

By combining the SAFiCF code~\cite{le_saficf} with the particle swarm optimization algorithm, we can fit the INS spectra by varying the CEF parameters. Calculations were also checked using the McPhase program~\cite{rotter_using_2004}. Fig.~\ref{fig:fit} plots the energy dependence of the INS intensities integrated within a momentum transfer of $1.2\sim2.2$ ($2.2\sim3.2$) \AA$^{-1}$  for the $E_i = 12$ (41)~meV data measured at $T = 5$ and 50 K, respectively. The calculated spectra are convoluted by a Gaussian function to account for the instrument resolution. The best fit is achieved with the set of CEF parameters shown in Table~\ref{tab:cef}. The tail-like broadening of the CEF excitations on the lower-$E$ side in Fig.~\ref{fig:fit}(a) might be related to the disorder at the Na sites as observed in our XRD refinement. The slight mismatch for the $E_i = 41$ meV data shown in Fig.~\ref{fig:fit}(b) and (d) might be due to the CEF-phonon hybridization as revealed from their $Q$-dependence together with imperfect descriptions of the background using only the polynomial terms.

The obtained CEF ground state wavefunctions for the Er$^{3+}$ ions are $|\pm\rangle = \pm0.123 |\pm11/2\rangle + 0.396|\pm5/2\rangle \pm 0.596|\mp1/2\rangle - 0.516|\mp 7/2\rangle \pm 0.453|\mp13/2\rangle$. The anisotropic $g$ factors are $g_\perp = 7.8$ in the $xy$ plane and $g_\parallel = 4.7$ along the $z$ direction, which sharply contrasts the Ising anisotopy observed in CdEr$_2X_4$ ($X=$ S, Se)~\cite{gao_dipolar_2018}. Such a difference originates from the different components of the CEF ground states: in CdEr$_2X_4$ ($X=$ S, Se), the ground states are dominated by the $|\pm15/2\rangle$ components, while in NaErS$_2$, the largest components are $|\pm1/2\rangle$. In the latter case, substantial quantum tunnelling can be expected, which will facilitate the development of the QSL state.

\begin{table}[b!]
\caption{The Wybourne CEF parameters (meV) for different rare earth ions.}
\label{tab:scale}
\centering
\begin{tabular}{lcccccc}
\toprule
 & $L_2^0$ & $L_4^0$ & $L_4^3$ & $L_6^0$ & $L_6^3$ & $L_6^6$ \\
 \hline
Nd$^{3+}$	& $-38.7$ & $-176.0$ & $-293.2\, $ & $91.4$ & $-0.4\,$ & $76.9$\\
Sm$^{3+}$ & $-33.8$ & $-136.7$ & $-227.7\, $ & $-$ & $-\,$ & $-$\\
Tb$^{3+}$	& $ -28.5$ & $ -99.8$ & $-166.3\, $ & $41.7$ & $ -0.2\,$ & $35.0$\\
Dy$^{3+}$	& $ -27.1$ & $-91.0$ & $-151.6\, $ & $36.8$ & $-0.1\,$ & $30.9$\\
Ho$^{3+}$	& $ -25.8$ & $-83.4$ & $-138.9\, $ & $32.7$ & $-0.1\,$ & $27.5$\\
Tm$^{3+}$	& $ -23.6$ & $-71.0$ & $ -118.2\, $ & $26.4$ & $-0.1\,$ & $22.2$\\
Yb$^{3+}$	& $ -22.6$ & $-65.9$ & $ -109.7\, $ & $23.9$ & $-0.1\,$ & $20.1$\\
\hline
\end{tabular}
\end{table}

\section{Discussions}\label{sec:disc}
The existence of the large $|J_z = \pm1/2\rangle$ components in the Er$^{3+}$ ground state wavefunctions can be understood through the point charge calculations. Assuming point charges of $-2e$ on the surrounding sulfur anion sites, the calculated CEF ground state doublet will be dominated by the $ |\pm15/2\rangle$ components with $g_\perp = 0$ and $g_\parallel = 17.8$. However, once the additional $+3e$ charges on the neighboring Er sites within the $ab$ plane are considered, the ground state wavefunctions will become  $|\pm\rangle = 0.004 |\pm13/2\rangle \pm 0.136|\pm7/2\rangle + 0.981|\pm1/2\rangle \mp 0.136|\mp 5/2\rangle + 0.002|\mp11/2\rangle$ with $g_\perp = 9.6$ and $g_\parallel = 1.2$, which is qualitatively similar to the results obtained from the INS spectra. Therefore, the electric charges beyond the ErS$_6$ octahedra play an important role in determining the Er$^{3+}$ ground state properties as for the Yb$^{3+}$ ions in YbMgGaO$_4$~\cite{li_crystalline_2017}.

The CEF parameters determined for Er$^{3+}$ in NaErS$_2$ can be scaled to other rare earth ions, thus providing basic knowledge on the CEF ground state in the sulfide delafossites. For this purpose, we first calculated the Hutchings CEF parameters for the Er$^{3+}$ ions~\cite{hutchings_point_1964}. The Hutchings CEF parameters depend only on the CEF environment and can be conveniently applied for systems with similar crystal structures~\cite{bertin_crystal_2012, gao_dipolar_2018}. Assuming the same Hutchings CEF parameters, Table~\ref{tab:scale} lists the corresponding Wybourne CEF parameters for different rare earth ions. Ce$^{3+}$ and Pr$^{3+}$ are omitted because the corresponding NaLnS$_2$ compounds do not crystallize in the $R\overline{3}m$  space group~\cite{liu_rare_2018}. For the Kramers ions, the CEF ground state doublet can be calculated as follows:
\begin{align}
\label{eq:scale_ground}
\text{Nd}^{3+}\ (J=9/2)\ |\pm\rangle &=0.346 |\pm 7/2\rangle
                \pm   0.122 |\pm 1/2\rangle \nonumber \\
              &+  0.930 |\mp 5/2 \rangle \text{,} \nonumber \\
\text{Sm}^{3+}\ (J=5/2)\ |\pm\rangle &=0.806 |\pm 5/2\rangle
                \pm  0.592 |\mp 1/2\rangle \nonumber \text{,}  \\
\text{Dy}^{3+}\ (J=15/2)\ |\pm\rangle &=0.318 |\pm13/2\rangle
              \mp 0.504 |\pm7/2\rangle \nonumber \\
              &+ 0.325 |\pm1/2\rangle 
              \pm0.536 |\mp5/2\rangle \nonumber \\ 
              &+ 0.501 |\mp 11/2\rangle \text{,} \nonumber \\               
\text{Yb}^{3+}\ (J=7/2)\ |\pm \rangle &= 0.484 |\pm 7/2\rangle
                \mp   0.525 |\pm 1/2\rangle \nonumber \\
               &-0.700 |\mp 5/2\rangle \text{.}            
\end{align}
Sizable $J_z = \pm 1/2$ components in the ground state doublet are predicted for all the Kramers ions, which supports the delafossites as candidate compounds for QSL states. Especially, in the case of Yb$^{3+}$, the scaled CEF parameters predict three excitations from the CEF ground state at 21.4, 30.0, and 55.8 meV with a cross section of 5.0, 4.0, and 0.1 barn, respectively. This calculation result is close to the experimental observation of two CEF transitions at 23 and 39 meV~\cite{baenitz_planar_2018}. Although the exact crystallographic structure and consequently the Hutchings CEF parameters depend on the rare earth ions, we expect the scaled CEF ground state wavefunction to be qualitatively correct~\cite{bertin_crystal_2012, gao_dipolar_2018}, which supports the ternary sufide delafossites as candidate compounds to realize the QSL state.

\section{Conclusions}\label{sec:disc}
INS experiments have been performed on the QSL candidate NaErS$_2$ to study the Er$^{3+}$ CEF transitions. The measured INS spectra can be fitted with the CEF Hamiltonian, which reveals the existence of large $J_z = \pm 1/2$ components in the ground state doublet that allows quantum fluctuations. Applying the fitted CEF parameters to other rare earth ions reveals that the $J_z = \pm 1/2$ components  also exist in the CEF ground states, supporting the rare-earth-based sulfide delafossites as candidate hosts for the QSL state.

\section{Acknowledgments}\label{sec:disc}
We thank T. Nakajima, M. Soda, L. Ding, and V. Kocsis for helpful discussions. Our inelastic neutron scattering experiment was performed at the Materials and Life Science Experimental Facility (MLF) of the Japan Proton Accelerator Research Complex (J-PARC) under the user program (Proposal No. 2019A0293). F.X. acknowledges the funding from the European Union's Horizon 2020 research and innovation program under the Marie Sk\l{}odowska-Curie grant agreement No 701647.
\vspace{5mm}

Added Note: \textit{The recent publication}~\cite{scheie_crystal_2020} \textit{on} KErS$_2$ \textit{and} CsErS$_2$ \textit{reveals similar CEF parameters and ground state wavefunctions as in our work.}


\begin{thebibliography}{55}%
\makeatletter
\providecommand \@ifxundefined [1]{%
 \@ifx{#1\undefined}
}%
\providecommand \@ifnum [1]{%
 \ifnum #1\expandafter \@firstoftwo
 \else \expandafter \@secondoftwo
 \fi
}%
\providecommand \@ifx [1]{%
 \ifx #1\expandafter \@firstoftwo
 \else \expandafter \@secondoftwo
 \fi
}%
\providecommand \natexlab [1]{#1}%
\providecommand \enquote  [1]{``#1''}%
\providecommand \bibnamefont  [1]{#1}%
\providecommand \bibfnamefont [1]{#1}%
\providecommand \citenamefont [1]{#1}%
\providecommand \href@noop [0]{\@secondoftwo}%
\providecommand \href [0]{\begingroup \@sanitize@url \@href}%
\providecommand \@href[1]{\@@startlink{#1}\@@href}%
\providecommand \@@href[1]{\endgroup#1\@@endlink}%
\providecommand \@sanitize@url [0]{\catcode `\\12\catcode `\$12\catcode
  `\&12\catcode `\#12\catcode `\^12\catcode `\_12\catcode `\%12\relax}%
\providecommand \@@startlink[1]{}%
\providecommand \@@endlink[0]{}%
\providecommand \url  [0]{\begingroup\@sanitize@url \@url }%
\providecommand \@url [1]{\endgroup\@href {#1}{\urlprefix }}%
\providecommand \urlprefix  [0]{URL }%
\providecommand \Eprint [0]{\href }%
\providecommand \doibase [0]{http://dx.doi.org/}%
\providecommand \selectlanguage [0]{\@gobble}%
\providecommand \bibinfo  [0]{\@secondoftwo}%
\providecommand \bibfield  [0]{\@secondoftwo}%
\providecommand \translation [1]{[#1]}%
\providecommand \BibitemOpen [0]{}%
\providecommand \bibitemStop [0]{}%
\providecommand \bibitemNoStop [0]{.\EOS\space}%
\providecommand \EOS [0]{\spacefactor3000\relax}%
\providecommand \BibitemShut  [1]{\csname bibitem#1\endcsname}%
\let\auto@bib@innerbib\@empty
\bibitem [{\citenamefont {Fazekas}\ and\ \citenamefont
  {Anderson}(1974)}]{fazekas_1974_on}%
  \BibitemOpen
  \bibfield  {author} {\bibinfo {author} {\bibfnamefont {P.}~\bibnamefont
  {Fazekas}}\ and\ \bibinfo {author} {\bibfnamefont {P.~W.}\ \bibnamefont
  {Anderson}},\ }\href {\doibase 10.1080/14786439808206568} {\bibfield
  {journal} {\bibinfo  {journal} {Philos. Mag.}\ }\textbf {\bibinfo {volume}
  {30}},\ \bibinfo {pages} {423} (\bibinfo {year} {1974})}\BibitemShut
  {NoStop}%
\bibitem [{\citenamefont {Giamarchi}(2003)}]{giamarchi_quantum_2003}%
  \BibitemOpen
  \bibfield  {author} {\bibinfo {author} {\bibfnamefont {T.}~\bibnamefont
  {Giamarchi}},\ }\href@noop {} {\emph {\bibinfo {title} {Quantum physics in
  one dimension}}}\ (\bibinfo  {publisher} {Oxford Science Publications,
  Oxford},\ \bibinfo {year} {2003})\BibitemShut {NoStop}%
\bibitem [{\citenamefont {Normand}(2009)}]{normand_frontiers_2009}%
  \BibitemOpen
  \bibfield  {author} {\bibinfo {author} {\bibfnamefont {B.}~\bibnamefont
  {Normand}},\ }\href {\doibase 10.1080/00107510902850361} {\bibfield
  {journal} {\bibinfo  {journal} {Contemp. Phys.}\ }\textbf {\bibinfo {volume}
  {50}},\ \bibinfo {pages} {533} (\bibinfo {year} {2009})}\BibitemShut
  {NoStop}%
\bibitem [{\citenamefont {Balents}(2010)}]{balents_spin_2010}%
  \BibitemOpen
  \bibfield  {author} {\bibinfo {author} {\bibfnamefont {L.}~\bibnamefont
  {Balents}},\ }\href {\doibase 10.1038/nature08917} {\bibfield  {journal}
  {\bibinfo  {journal} {Nature}\ }\textbf {\bibinfo {volume} {464}},\ \bibinfo
  {pages} {199} (\bibinfo {year} {2010})}\BibitemShut {NoStop}%
\bibitem [{\citenamefont {Savary}\ and\ \citenamefont
  {Balents}(2016)}]{savary_quantum_2016}%
  \BibitemOpen
  \bibfield  {author} {\bibinfo {author} {\bibfnamefont {L.}~\bibnamefont
  {Savary}}\ and\ \bibinfo {author} {\bibfnamefont {L.}~\bibnamefont
  {Balents}},\ }\href {\doibase 10.1088/0034-4885/80/1/016502} {\bibfield
  {journal} {\bibinfo  {journal} {Rep. Prog. Phys.}\ }\textbf {\bibinfo
  {volume} {80}},\ \bibinfo {pages} {016502} (\bibinfo {year}
  {2016})}\BibitemShut {NoStop}%
\bibitem [{\citenamefont {Zhou}\ \emph {et~al.}(2017)\citenamefont {Zhou},
  \citenamefont {Kanoda},\ and\ \citenamefont {Ng}}]{zhou_quantum_2017}%
  \BibitemOpen
  \bibfield  {author} {\bibinfo {author} {\bibfnamefont {Y.}~\bibnamefont
  {Zhou}}, \bibinfo {author} {\bibfnamefont {K.}~\bibnamefont {Kanoda}}, \ and\
  \bibinfo {author} {\bibfnamefont {T.-K.}\ \bibnamefont {Ng}},\ }\href
  {\doibase 10.1103/RevModPhys.89.025003} {\bibfield  {journal} {\bibinfo
  {journal} {Rev. Mod. Phys.}\ }\textbf {\bibinfo {volume} {89}},\ \bibinfo
  {pages} {025003} (\bibinfo {year} {2017})}\BibitemShut {NoStop}%
\bibitem [{\citenamefont {Kitaev}(2006)}]{kitaev_anyons_2006}%
  \BibitemOpen
  \bibfield  {author} {\bibinfo {author} {\bibfnamefont {A.}~\bibnamefont
  {Kitaev}},\ }\href {\doibase 10.1016/j.aop.2005.10.005} {\bibfield  {journal}
  {\bibinfo  {journal} {Ann. Phys.}\ }\textbf {\bibinfo {volume} {321}},\
  \bibinfo {pages} {2} (\bibinfo {year} {2006})}\BibitemShut {NoStop}%
\bibitem [{\citenamefont {Nayak}\ \emph {et~al.}(2008)\citenamefont {Nayak},
  \citenamefont {Simon}, \citenamefont {Stern}, \citenamefont {Freedman},\ and\
  \citenamefont {Das~Sarma}}]{nayak_non_2008}%
  \BibitemOpen
  \bibfield  {author} {\bibinfo {author} {\bibfnamefont {C.}~\bibnamefont
  {Nayak}}, \bibinfo {author} {\bibfnamefont {S.~H.}\ \bibnamefont {Simon}},
  \bibinfo {author} {\bibfnamefont {A.}~\bibnamefont {Stern}}, \bibinfo
  {author} {\bibfnamefont {M.}~\bibnamefont {Freedman}}, \ and\ \bibinfo
  {author} {\bibfnamefont {S.}~\bibnamefont {Das~Sarma}},\ }\href {\doibase
  10.1103/RevModPhys.80.1083} {\bibfield  {journal} {\bibinfo  {journal} {Rev.
  Mod. Phys.}\ }\textbf {\bibinfo {volume} {80}},\ \bibinfo {pages} {1083}
  (\bibinfo {year} {2008})}\BibitemShut {NoStop}%
\bibitem [{\citenamefont {Shimizu}\ \emph {et~al.}(2003)\citenamefont
  {Shimizu}, \citenamefont {Miyagawa}, \citenamefont {Kanoda}, \citenamefont
  {Maesato},\ and\ \citenamefont {Saito}}]{shimizu_spin_2003}%
  \BibitemOpen
  \bibfield  {author} {\bibinfo {author} {\bibfnamefont {Y.}~\bibnamefont
  {Shimizu}}, \bibinfo {author} {\bibfnamefont {K.}~\bibnamefont {Miyagawa}},
  \bibinfo {author} {\bibfnamefont {K.}~\bibnamefont {Kanoda}}, \bibinfo
  {author} {\bibfnamefont {M.}~\bibnamefont {Maesato}}, \ and\ \bibinfo
  {author} {\bibfnamefont {G.}~\bibnamefont {Saito}},\ }\href {\doibase
  10.1103/PhysRevLett.91.107001} {\bibfield  {journal} {\bibinfo  {journal}
  {Phys. Rev. Lett.}\ }\textbf {\bibinfo {volume} {91}},\ \bibinfo {pages}
  {107001} (\bibinfo {year} {2003})}\BibitemShut {NoStop}%
\bibitem [{\citenamefont {Mendels}\ \emph {et~al.}(2007)\citenamefont
  {Mendels}, \citenamefont {Bert}, \citenamefont {de~Vries}, \citenamefont
  {Olariu}, \citenamefont {Harrison}, \citenamefont {Duc}, \citenamefont
  {Trombe}, \citenamefont {Lord}, \citenamefont {Amato},\ and\ \citenamefont
  {Baines}}]{mendels_quantum_2007}%
  \BibitemOpen
  \bibfield  {author} {\bibinfo {author} {\bibfnamefont {P.}~\bibnamefont
  {Mendels}}, \bibinfo {author} {\bibfnamefont {F.}~\bibnamefont {Bert}},
  \bibinfo {author} {\bibfnamefont {M.~A.}\ \bibnamefont {de~Vries}}, \bibinfo
  {author} {\bibfnamefont {A.}~\bibnamefont {Olariu}}, \bibinfo {author}
  {\bibfnamefont {A.}~\bibnamefont {Harrison}}, \bibinfo {author}
  {\bibfnamefont {F.}~\bibnamefont {Duc}}, \bibinfo {author} {\bibfnamefont
  {J.~C.}\ \bibnamefont {Trombe}}, \bibinfo {author} {\bibfnamefont {J.~S.}\
  \bibnamefont {Lord}}, \bibinfo {author} {\bibfnamefont {A.}~\bibnamefont
  {Amato}}, \ and\ \bibinfo {author} {\bibfnamefont {C.}~\bibnamefont
  {Baines}},\ }\href {\doibase 10.1103/PhysRevLett.98.077204} {\bibfield
  {journal} {\bibinfo  {journal} {Phys. Rev. Lett.}\ }\textbf {\bibinfo
  {volume} {98}},\ \bibinfo {pages} {077204} (\bibinfo {year}
  {2007})}\BibitemShut {NoStop}%
\bibitem [{\citenamefont {Han}\ \emph {et~al.}(2012)\citenamefont {Han},
  \citenamefont {Helton}, \citenamefont {Chu}, \citenamefont {Nocera},
  \citenamefont {Rodriguez-Rivera}, \citenamefont {Broholm},\ and\
  \citenamefont {Lee}}]{han_fractionalized_2012}%
  \BibitemOpen
  \bibfield  {author} {\bibinfo {author} {\bibfnamefont {T.-H.}\ \bibnamefont
  {Han}}, \bibinfo {author} {\bibfnamefont {J.~S.}\ \bibnamefont {Helton}},
  \bibinfo {author} {\bibfnamefont {S.}~\bibnamefont {Chu}}, \bibinfo {author}
  {\bibfnamefont {D.~G.}\ \bibnamefont {Nocera}}, \bibinfo {author}
  {\bibfnamefont {J.~A.}\ \bibnamefont {Rodriguez-Rivera}}, \bibinfo {author}
  {\bibfnamefont {C.}~\bibnamefont {Broholm}}, \ and\ \bibinfo {author}
  {\bibfnamefont {Y.~S.}\ \bibnamefont {Lee}},\ }\href {\doibase
  10.1038/nature11659} {\bibfield  {journal} {\bibinfo  {journal} {Nature}\
  }\textbf {\bibinfo {volume} {492}},\ \bibinfo {pages} {406} (\bibinfo {year}
  {2012})}\BibitemShut {NoStop}%
\bibitem [{\citenamefont {Fu}\ \emph {et~al.}(2015)\citenamefont {Fu},
  \citenamefont {Imai}, \citenamefont {Han},\ and\ \citenamefont
  {Lee}}]{fu_evidence_2015}%
  \BibitemOpen
  \bibfield  {author} {\bibinfo {author} {\bibfnamefont {M.}~\bibnamefont
  {Fu}}, \bibinfo {author} {\bibfnamefont {T.}~\bibnamefont {Imai}}, \bibinfo
  {author} {\bibfnamefont {T.-H.}\ \bibnamefont {Han}}, \ and\ \bibinfo
  {author} {\bibfnamefont {Y.~S.}\ \bibnamefont {Lee}},\ }\href {\doibase
  10.1126/science.aab2120} {\bibfield  {journal} {\bibinfo  {journal}
  {Science}\ }\textbf {\bibinfo {volume} {350}},\ \bibinfo {pages} {655}
  (\bibinfo {year} {2015})}\BibitemShut {NoStop}%
\bibitem [{\citenamefont {Rau}\ and\ \citenamefont
  {Gingras}(2015)}]{rau_magnitude_2015}%
  \BibitemOpen
  \bibfield  {author} {\bibinfo {author} {\bibfnamefont {J.~G.}\ \bibnamefont
  {Rau}}\ and\ \bibinfo {author} {\bibfnamefont {M.~J.~P.}\ \bibnamefont
  {Gingras}},\ }\href {\doibase 10.1103/PhysRevB.92.144417} {\bibfield
  {journal} {\bibinfo  {journal} {Phys. Rev. B}\ }\textbf {\bibinfo {volume}
  {92}},\ \bibinfo {pages} {144417} (\bibinfo {year} {2015})}\BibitemShut
  {NoStop}%
\bibitem [{\citenamefont {Rau}\ and\ \citenamefont
  {Gingras}(2018)}]{rau_frustration_2018}%
  \BibitemOpen
  \bibfield  {author} {\bibinfo {author} {\bibfnamefont {J.~G.}\ \bibnamefont
  {Rau}}\ and\ \bibinfo {author} {\bibfnamefont {M.~J.~P.}\ \bibnamefont
  {Gingras}},\ }\href {\doibase 10.1103/PhysRevB.98.054408} {\bibfield
  {journal} {\bibinfo  {journal} {Phys. Rev. B}\ }\textbf {\bibinfo {volume}
  {98}},\ \bibinfo {pages} {054408} (\bibinfo {year} {2018})}\BibitemShut
  {NoStop}%
\bibitem [{\citenamefont {Hermele}\ \emph {et~al.}(2004)\citenamefont
  {Hermele}, \citenamefont {Fisher},\ and\ \citenamefont
  {Balents}}]{hermele_pyrochlore_2004}%
  \BibitemOpen
  \bibfield  {author} {\bibinfo {author} {\bibfnamefont {M.}~\bibnamefont
  {Hermele}}, \bibinfo {author} {\bibfnamefont {M.~P.~A.}\ \bibnamefont
  {Fisher}}, \ and\ \bibinfo {author} {\bibfnamefont {L.}~\bibnamefont
  {Balents}},\ }\href {\doibase 10.1103/PhysRevB.69.064404} {\bibfield
  {journal} {\bibinfo  {journal} {Phys. Rev. B}\ }\textbf {\bibinfo {volume}
  {69}},\ \bibinfo {pages} {64404} (\bibinfo {year} {2004})}\BibitemShut
  {NoStop}%
\bibitem [{\citenamefont {Gingras}\ and\ \citenamefont
  {McClarty}(2014)}]{gingras_quantum_2014}%
  \BibitemOpen
  \bibfield  {author} {\bibinfo {author} {\bibfnamefont {M.~J.~P.}\
  \bibnamefont {Gingras}}\ and\ \bibinfo {author} {\bibfnamefont {P.~A.}\
  \bibnamefont {McClarty}},\ }\href {\doibase 10.1088/0034-4885/77/5/056501}
  {\bibfield  {journal} {\bibinfo  {journal} {Rep. Prog. Phys.}\ }\textbf
  {\bibinfo {volume} {77}},\ \bibinfo {pages} {056501} (\bibinfo {year}
  {2014})}\BibitemShut {NoStop}%
\bibitem [{\citenamefont {Bramwell}\ and\ \citenamefont
  {Gingras}(2001)}]{bramwell_spin_2001}%
  \BibitemOpen
  \bibfield  {author} {\bibinfo {author} {\bibfnamefont {S.~T.}\ \bibnamefont
  {Bramwell}}\ and\ \bibinfo {author} {\bibfnamefont {M.~J.~P.}\ \bibnamefont
  {Gingras}},\ }\href {\doibase 10.1126/science.1064761} {\bibfield  {journal}
  {\bibinfo  {journal} {Science}\ }\textbf {\bibinfo {volume} {294}},\ \bibinfo
  {pages} {1495} (\bibinfo {year} {2001})}\BibitemShut {NoStop}%
\bibitem [{\citenamefont {Fennell}\ \emph {et~al.}(2009)\citenamefont
  {Fennell}, \citenamefont {Deen}, \citenamefont {Wildes}, \citenamefont
  {Schmalzl}, \citenamefont {Prabhakaran}, \citenamefont {Boothroyd},
  \citenamefont {Aldus}, \citenamefont {McMorrow},\ and\ \citenamefont
  {Bramwell}}]{fennell_magnetic_2009}%
  \BibitemOpen
  \bibfield  {author} {\bibinfo {author} {\bibfnamefont {T.}~\bibnamefont
  {Fennell}}, \bibinfo {author} {\bibfnamefont {P.~P.}\ \bibnamefont {Deen}},
  \bibinfo {author} {\bibfnamefont {A.~R.}\ \bibnamefont {Wildes}}, \bibinfo
  {author} {\bibfnamefont {K.}~\bibnamefont {Schmalzl}}, \bibinfo {author}
  {\bibfnamefont {D.}~\bibnamefont {Prabhakaran}}, \bibinfo {author}
  {\bibfnamefont {A.~T.}\ \bibnamefont {Boothroyd}}, \bibinfo {author}
  {\bibfnamefont {R.~J.}\ \bibnamefont {Aldus}}, \bibinfo {author}
  {\bibfnamefont {D.~F.}\ \bibnamefont {McMorrow}}, \ and\ \bibinfo {author}
  {\bibfnamefont {S.~T.}\ \bibnamefont {Bramwell}},\ }\href {\doibase
  10.1126/science.1177582} {\bibfield  {journal} {\bibinfo  {journal}
  {Science}\ }\textbf {\bibinfo {volume} {326}},\ \bibinfo {pages} {415}
  (\bibinfo {year} {2009})}\BibitemShut {NoStop}%
\bibitem [{\citenamefont {Fennell}\ \emph {et~al.}(2012)\citenamefont
  {Fennell}, \citenamefont {Kenzelmann}, \citenamefont {Roessli}, \citenamefont
  {Haas},\ and\ \citenamefont {Cava}}]{fennell_power_2012}%
  \BibitemOpen
  \bibfield  {author} {\bibinfo {author} {\bibfnamefont {T.}~\bibnamefont
  {Fennell}}, \bibinfo {author} {\bibfnamefont {M.}~\bibnamefont {Kenzelmann}},
  \bibinfo {author} {\bibfnamefont {B.}~\bibnamefont {Roessli}}, \bibinfo
  {author} {\bibfnamefont {M.~K.}\ \bibnamefont {Haas}}, \ and\ \bibinfo
  {author} {\bibfnamefont {R.~J.}\ \bibnamefont {Cava}},\ }\href {\doibase
  10.1103/PhysRevLett.109.017201} {\bibfield  {journal} {\bibinfo  {journal}
  {Phys. Rev. Lett.}\ }\textbf {\bibinfo {volume} {109}},\ \bibinfo {pages}
  {17201} (\bibinfo {year} {2012})}\BibitemShut {NoStop}%
\bibitem [{\citenamefont {Chang}\ \emph {et~al.}(2012)\citenamefont {Chang},
  \citenamefont {Onoda}, \citenamefont {Su}, \citenamefont {Kao}, \citenamefont
  {Tsuei}, \citenamefont {Yasui}, \citenamefont {Kakurai},\ and\ \citenamefont
  {Lees}}]{chang_higgs_2012}%
  \BibitemOpen
  \bibfield  {author} {\bibinfo {author} {\bibfnamefont {L.-J.}\ \bibnamefont
  {Chang}}, \bibinfo {author} {\bibfnamefont {S.}~\bibnamefont {Onoda}},
  \bibinfo {author} {\bibfnamefont {Y.}~\bibnamefont {Su}}, \bibinfo {author}
  {\bibfnamefont {Y.-J.}\ \bibnamefont {Kao}}, \bibinfo {author} {\bibfnamefont
  {K.-D.}\ \bibnamefont {Tsuei}}, \bibinfo {author} {\bibfnamefont
  {Y.}~\bibnamefont {Yasui}}, \bibinfo {author} {\bibfnamefont
  {K.}~\bibnamefont {Kakurai}}, \ and\ \bibinfo {author} {\bibfnamefont
  {M.~R.}\ \bibnamefont {Lees}},\ }\href {\doibase 10.1038/ncomms1989}
  {\bibfield  {journal} {\bibinfo  {journal} {Nat. Commun.}\ }\textbf {\bibinfo
  {volume} {3}},\ \bibinfo {pages} {992} (\bibinfo {year} {2012})}\BibitemShut
  {NoStop}%
\bibitem [{\citenamefont {{Sibille}}\ \emph {et~al.}(2018)\citenamefont
  {{Sibille}}, \citenamefont {{Gauthier}}, \citenamefont {{Yan}}, \citenamefont
  {{Ciomaga Hatnean}}, \citenamefont {{Ollivier}}, \citenamefont {{Winn}},
  \citenamefont {{Filges}}, \citenamefont {{Balakrishnan}}, \citenamefont
  {{Kenzelmann}}, \citenamefont {{Shannon}},\ and\ \citenamefont
  {{Fennell}}}]{sibille_experimental_2018}%
  \BibitemOpen
  \bibfield  {author} {\bibinfo {author} {\bibfnamefont {R.}~\bibnamefont
  {{Sibille}}}, \bibinfo {author} {\bibfnamefont {N.}~\bibnamefont
  {{Gauthier}}}, \bibinfo {author} {\bibfnamefont {H.}~\bibnamefont {{Yan}}},
  \bibinfo {author} {\bibfnamefont {M.}~\bibnamefont {{Ciomaga Hatnean}}},
  \bibinfo {author} {\bibfnamefont {J.}~\bibnamefont {{Ollivier}}}, \bibinfo
  {author} {\bibfnamefont {B.}~\bibnamefont {{Winn}}}, \bibinfo {author}
  {\bibfnamefont {U.}~\bibnamefont {{Filges}}}, \bibinfo {author}
  {\bibfnamefont {G.}~\bibnamefont {{Balakrishnan}}}, \bibinfo {author}
  {\bibfnamefont {M.}~\bibnamefont {{Kenzelmann}}}, \bibinfo {author}
  {\bibfnamefont {N.}~\bibnamefont {{Shannon}}}, \ and\ \bibinfo {author}
  {\bibfnamefont {T.}~\bibnamefont {{Fennell}}},\ }\href {\doibase
  10.1038/s41567-018-0116-x} {\bibfield  {journal} {\bibinfo  {journal} {Nat.
  Phys.}\ }\textbf {\bibinfo {volume} {14}},\ \bibinfo {pages} {711} (\bibinfo
  {year} {2018})}\BibitemShut {NoStop}%
\bibitem [{\citenamefont {Benton}\ \emph {et~al.}(2012)\citenamefont {Benton},
  \citenamefont {Sikora},\ and\ \citenamefont {Shannon}}]{benton_seeing_2012}%
  \BibitemOpen
  \bibfield  {author} {\bibinfo {author} {\bibfnamefont {O.}~\bibnamefont
  {Benton}}, \bibinfo {author} {\bibfnamefont {O.}~\bibnamefont {Sikora}}, \
  and\ \bibinfo {author} {\bibfnamefont {N.}~\bibnamefont {Shannon}},\ }\href
  {\doibase 10.1103/PhysRevB.86.075154} {\bibfield  {journal} {\bibinfo
  {journal} {Phys. Rev. B}\ }\textbf {\bibinfo {volume} {86}},\ \bibinfo
  {pages} {075154} (\bibinfo {year} {2012})}\BibitemShut {NoStop}%
\bibitem [{\citenamefont {Li}\ \emph {et~al.}(2016)\citenamefont {Li},
  \citenamefont {Wang},\ and\ \citenamefont {Chen}}]{li_anisotropic_2016}%
  \BibitemOpen
  \bibfield  {author} {\bibinfo {author} {\bibfnamefont {Y.-D.}\ \bibnamefont
  {Li}}, \bibinfo {author} {\bibfnamefont {X.}~\bibnamefont {Wang}}, \ and\
  \bibinfo {author} {\bibfnamefont {G.}~\bibnamefont {Chen}},\ }\href {\doibase
  10.1103/PhysRevB.94.035107} {\bibfield  {journal} {\bibinfo  {journal} {Phys.
  Rev. B}\ }\textbf {\bibinfo {volume} {94}},\ \bibinfo {pages} {035107}
  (\bibinfo {year} {2016})}\BibitemShut {NoStop}%
\bibitem [{\citenamefont {Li}\ \emph {et~al.}(2015)\citenamefont {Li},
  \citenamefont {Chen}, \citenamefont {Tong}, \citenamefont {Pi}, \citenamefont
  {Liu}, \citenamefont {Yang}, \citenamefont {Wang},\ and\ \citenamefont
  {Zhang}}]{li_rare_2015}%
  \BibitemOpen
  \bibfield  {author} {\bibinfo {author} {\bibfnamefont {Y.}~\bibnamefont
  {Li}}, \bibinfo {author} {\bibfnamefont {G.}~\bibnamefont {Chen}}, \bibinfo
  {author} {\bibfnamefont {W.}~\bibnamefont {Tong}}, \bibinfo {author}
  {\bibfnamefont {L.}~\bibnamefont {Pi}}, \bibinfo {author} {\bibfnamefont
  {J.}~\bibnamefont {Liu}}, \bibinfo {author} {\bibfnamefont {Z.}~\bibnamefont
  {Yang}}, \bibinfo {author} {\bibfnamefont {X.}~\bibnamefont {Wang}}, \ and\
  \bibinfo {author} {\bibfnamefont {Q.}~\bibnamefont {Zhang}},\ }\href
  {\doibase 10.1103/PhysRevLett.115.167203} {\bibfield  {journal} {\bibinfo
  {journal} {Phys. Rev. Lett.}\ }\textbf {\bibinfo {volume} {115}},\ \bibinfo
  {pages} {167203} (\bibinfo {year} {2015})}\BibitemShut {NoStop}%
\bibitem [{\citenamefont {Paddison}\ \emph {et~al.}(2016)\citenamefont
  {Paddison}, \citenamefont {Daum}, \citenamefont {Dun}, \citenamefont
  {Ehlers}, \citenamefont {Liu}, \citenamefont {Stone}, \citenamefont {Zhou},\
  and\ \citenamefont {Mourigal}}]{paddison_continuous_2016}%
  \BibitemOpen
  \bibfield  {author} {\bibinfo {author} {\bibfnamefont {J.~A.~M.}\
  \bibnamefont {Paddison}}, \bibinfo {author} {\bibfnamefont {M.}~\bibnamefont
  {Daum}}, \bibinfo {author} {\bibfnamefont {Z.}~\bibnamefont {Dun}}, \bibinfo
  {author} {\bibfnamefont {G.}~\bibnamefont {Ehlers}}, \bibinfo {author}
  {\bibfnamefont {Y.}~\bibnamefont {Liu}}, \bibinfo {author} {\bibfnamefont
  {M.~B.}\ \bibnamefont {Stone}}, \bibinfo {author} {\bibfnamefont
  {H.}~\bibnamefont {Zhou}}, \ and\ \bibinfo {author} {\bibfnamefont
  {M.}~\bibnamefont {Mourigal}},\ }\href {\doibase 10.1038/nphys3971}
  {\bibfield  {journal} {\bibinfo  {journal} {Nat. Phys.}\ }\textbf {\bibinfo
  {volume} {13}},\ \bibinfo {pages} {117} (\bibinfo {year} {2016})}\BibitemShut
  {NoStop}%
\bibitem [{\citenamefont {Shen}\ \emph {et~al.}(2016)\citenamefont {Shen},
  \citenamefont {Li}, \citenamefont {Wo}, \citenamefont {Li}, \citenamefont
  {Shen}, \citenamefont {Pan}, \citenamefont {Wang}, \citenamefont {Walker},
  \citenamefont {Steffens}, \citenamefont {Boehm}, \citenamefont {Hao},
  \citenamefont {Quintero-Castro}, \citenamefont {Harriger}, \citenamefont
  {Frontzek}, \citenamefont {Hao}, \citenamefont {Meng}, \citenamefont {Zhang},
  \citenamefont {Chen},\ and\ \citenamefont {Zhao}}]{shen_evidence_2016}%
  \BibitemOpen
  \bibfield  {author} {\bibinfo {author} {\bibfnamefont {Y.}~\bibnamefont
  {Shen}}, \bibinfo {author} {\bibfnamefont {Y.-D.}\ \bibnamefont {Li}},
  \bibinfo {author} {\bibfnamefont {H.}~\bibnamefont {Wo}}, \bibinfo {author}
  {\bibfnamefont {Y.}~\bibnamefont {Li}}, \bibinfo {author} {\bibfnamefont
  {S.}~\bibnamefont {Shen}}, \bibinfo {author} {\bibfnamefont {B.}~\bibnamefont
  {Pan}}, \bibinfo {author} {\bibfnamefont {Q.}~\bibnamefont {Wang}}, \bibinfo
  {author} {\bibfnamefont {H.~C.}\ \bibnamefont {Walker}}, \bibinfo {author}
  {\bibfnamefont {P.}~\bibnamefont {Steffens}}, \bibinfo {author}
  {\bibfnamefont {M.}~\bibnamefont {Boehm}}, \bibinfo {author} {\bibfnamefont
  {Y.}~\bibnamefont {Hao}}, \bibinfo {author} {\bibfnamefont {D.~L.}\
  \bibnamefont {Quintero-Castro}}, \bibinfo {author} {\bibfnamefont {L.~W.}\
  \bibnamefont {Harriger}}, \bibinfo {author} {\bibfnamefont {M.~D.}\
  \bibnamefont {Frontzek}}, \bibinfo {author} {\bibfnamefont {L.}~\bibnamefont
  {Hao}}, \bibinfo {author} {\bibfnamefont {S.}~\bibnamefont {Meng}}, \bibinfo
  {author} {\bibfnamefont {Q.}~\bibnamefont {Zhang}}, \bibinfo {author}
  {\bibfnamefont {G.}~\bibnamefont {Chen}}, \ and\ \bibinfo {author}
  {\bibfnamefont {J.}~\bibnamefont {Zhao}},\ }\href {\doibase
  10.1038/nature20614} {\bibfield  {journal} {\bibinfo  {journal} {Nature}\
  }\textbf {\bibinfo {volume} {540}},\ \bibinfo {pages} {559} (\bibinfo {year}
  {2016})}\BibitemShut {NoStop}%
\bibitem [{\citenamefont {Xu}\ \emph {et~al.}(2016)\citenamefont {Xu},
  \citenamefont {Zhang}, \citenamefont {Li}, \citenamefont {Yu}, \citenamefont
  {Hong}, \citenamefont {Zhang},\ and\ \citenamefont {Li}}]{xu_absence_2016}%
  \BibitemOpen
  \bibfield  {author} {\bibinfo {author} {\bibfnamefont {Y.}~\bibnamefont
  {Xu}}, \bibinfo {author} {\bibfnamefont {J.}~\bibnamefont {Zhang}}, \bibinfo
  {author} {\bibfnamefont {Y.~S.}\ \bibnamefont {Li}}, \bibinfo {author}
  {\bibfnamefont {Y.~J.}\ \bibnamefont {Yu}}, \bibinfo {author} {\bibfnamefont
  {X.~C.}\ \bibnamefont {Hong}}, \bibinfo {author} {\bibfnamefont {Q.~M.}\
  \bibnamefont {Zhang}}, \ and\ \bibinfo {author} {\bibfnamefont {S.~Y.}\
  \bibnamefont {Li}},\ }\href {\doibase 10.1103/PhysRevLett.117.267202}
  {\bibfield  {journal} {\bibinfo  {journal} {Phys. Rev. Lett.}\ }\textbf
  {\bibinfo {volume} {117}},\ \bibinfo {pages} {267202} (\bibinfo {year}
  {2016})}\BibitemShut {NoStop}%
\bibitem [{\citenamefont {Zhu}\ \emph {et~al.}(2017)\citenamefont {Zhu},
  \citenamefont {Maksimov}, \citenamefont {White},\ and\ \citenamefont
  {Chernyshev}}]{zhu_disorder_2017}%
  \BibitemOpen
  \bibfield  {author} {\bibinfo {author} {\bibfnamefont {Z.}~\bibnamefont
  {Zhu}}, \bibinfo {author} {\bibfnamefont {P.~A.}\ \bibnamefont {Maksimov}},
  \bibinfo {author} {\bibfnamefont {S.~R.}\ \bibnamefont {White}}, \ and\
  \bibinfo {author} {\bibfnamefont {A.~L.}\ \bibnamefont {Chernyshev}},\ }\href
  {\doibase 10.1103/PhysRevLett.119.157201} {\bibfield  {journal} {\bibinfo
  {journal} {Phys. Rev. Lett.}\ }\textbf {\bibinfo {volume} {119}},\ \bibinfo
  {pages} {157201} (\bibinfo {year} {2017})}\BibitemShut {NoStop}%
\bibitem [{\citenamefont {Kimchi}\ \emph {et~al.}(2018)\citenamefont {Kimchi},
  \citenamefont {Nahum},\ and\ \citenamefont {Senthil}}]{kimchi_valence_2018}%
  \BibitemOpen
  \bibfield  {author} {\bibinfo {author} {\bibfnamefont {I.}~\bibnamefont
  {Kimchi}}, \bibinfo {author} {\bibfnamefont {A.}~\bibnamefont {Nahum}}, \
  and\ \bibinfo {author} {\bibfnamefont {T.}~\bibnamefont {Senthil}},\ }\href
  {\doibase 10.1103/PhysRevX.8.031028} {\bibfield  {journal} {\bibinfo
  {journal} {Phys. Rev. X}\ }\textbf {\bibinfo {volume} {8}},\ \bibinfo {pages}
  {031028} (\bibinfo {year} {2018})}\BibitemShut {NoStop}%
\bibitem [{\citenamefont {Schleid}\ and\ \citenamefont
  {Lissner}(1993)}]{schleid_single_1993}%
  \BibitemOpen
  \bibfield  {author} {\bibinfo {author} {\bibfnamefont {T.}~\bibnamefont
  {Schleid}}\ and\ \bibinfo {author} {\bibfnamefont {F.}~\bibnamefont
  {Lissner}},\ }\href@noop {} {\bibfield  {journal} {\bibinfo  {journal} {Eur.
  J. Inorg. Chem.}\ }\textbf {\bibinfo {volume} {30}},\ \bibinfo {pages} {829}
  (\bibinfo {year} {1993})}\BibitemShut {NoStop}%
\bibitem [{\citenamefont {{Liu}}\ \emph {et~al.}(2018)\citenamefont {{Liu}},
  \citenamefont {{Zhang}}, \citenamefont {{Ji}}, \citenamefont {{Liu}},
  \citenamefont {{Li}}, \citenamefont {{Wang}}, \citenamefont {{Lei}},
  \citenamefont {{Chen}},\ and\ \citenamefont {{Zhang}}}]{liu_rare_2018}%
  \BibitemOpen
  \bibfield  {author} {\bibinfo {author} {\bibfnamefont {W.}~\bibnamefont
  {{Liu}}}, \bibinfo {author} {\bibfnamefont {Z.}~\bibnamefont {{Zhang}}},
  \bibinfo {author} {\bibfnamefont {J.}~\bibnamefont {{Ji}}}, \bibinfo {author}
  {\bibfnamefont {Y.}~\bibnamefont {{Liu}}}, \bibinfo {author} {\bibfnamefont
  {J.}~\bibnamefont {{Li}}}, \bibinfo {author} {\bibfnamefont {X.}~\bibnamefont
  {{Wang}}}, \bibinfo {author} {\bibfnamefont {H.}~\bibnamefont {{Lei}}},
  \bibinfo {author} {\bibfnamefont {G.}~\bibnamefont {{Chen}}}, \ and\ \bibinfo
  {author} {\bibfnamefont {Q.}~\bibnamefont {{Zhang}}},\ }\href {\doibase
  10.1088/0256-307X/35/11/117501} {\bibfield  {journal} {\bibinfo  {journal}
  {Chin. Phys. Lett.}\ }\textbf {\bibinfo {volume} {35}},\ \bibinfo {eid}
  {117501} (\bibinfo {year} {2018})}\BibitemShut {NoStop}%
\bibitem [{\citenamefont {Hashimoto}\ \emph {et~al.}(2003)\citenamefont
  {Hashimoto}, \citenamefont {Wakeshima},\ and\ \citenamefont
  {Hinatsu}}]{hashimoto_magnetic_2003}%
  \BibitemOpen
  \bibfield  {author} {\bibinfo {author} {\bibfnamefont {Y.}~\bibnamefont
  {Hashimoto}}, \bibinfo {author} {\bibfnamefont {M.}~\bibnamefont
  {Wakeshima}}, \ and\ \bibinfo {author} {\bibfnamefont {Y.}~\bibnamefont
  {Hinatsu}},\ }\href {\doibase 10.1016/j.jssc.2003.08.001} {\bibfield
  {journal} {\bibinfo  {journal} {J. Solid State Chem.}\ }\textbf {\bibinfo
  {volume} {176}},\ \bibinfo {pages} {266} (\bibinfo {year}
  {2003})}\BibitemShut {NoStop}%
\bibitem [{\citenamefont {Dong}\ \emph {et~al.}(2008)\citenamefont {Dong},
  \citenamefont {Doi},\ and\ \citenamefont {Hinatsu}}]{dong_structure_2008}%
  \BibitemOpen
  \bibfield  {author} {\bibinfo {author} {\bibfnamefont {B.}~\bibnamefont
  {Dong}}, \bibinfo {author} {\bibfnamefont {Y.}~\bibnamefont {Doi}}, \ and\
  \bibinfo {author} {\bibfnamefont {Y.}~\bibnamefont {Hinatsu}},\ }\href
  {\doibase 10.1016/j.jallcom.2006.11.053} {\bibfield  {journal} {\bibinfo
  {journal} {J. Alloys Compd.}\ }\textbf {\bibinfo {volume} {453}},\ \bibinfo
  {pages} {282} (\bibinfo {year} {2008})}\BibitemShut {NoStop}%
\bibitem [{\citenamefont {Miyasaka}\ \emph {et~al.}(2009)\citenamefont
  {Miyasaka}, \citenamefont {Doi},\ and\ \citenamefont
  {Hinatsu}}]{miyasaka_synthesis_2009}%
  \BibitemOpen
  \bibfield  {author} {\bibinfo {author} {\bibfnamefont {N.}~\bibnamefont
  {Miyasaka}}, \bibinfo {author} {\bibfnamefont {Y.}~\bibnamefont {Doi}}, \
  and\ \bibinfo {author} {\bibfnamefont {Y.}~\bibnamefont {Hinatsu}},\ }\href
  {\doibase 10.1016/j.jssc.2009.05.035} {\bibfield  {journal} {\bibinfo
  {journal} {J. Solid State Chem.}\ }\textbf {\bibinfo {volume} {182}},\
  \bibinfo {pages} {2104} (\bibinfo {year} {2009})}\BibitemShut {NoStop}%
\bibitem [{\citenamefont {Baenitz}\ \emph {et~al.}(2018)\citenamefont
  {Baenitz}, \citenamefont {Schlender}, \citenamefont {Sichelschmidt},
  \citenamefont {Onykiienko}, \citenamefont {Zangeneh}, \citenamefont
  {Ranjith}, \citenamefont {Sarkar}, \citenamefont {Hozoi}, \citenamefont
  {Walker}, \citenamefont {Orain}, \citenamefont {Yasuoka}, \citenamefont
  {van~den Brink}, \citenamefont {Klauss}, \citenamefont {Inosov},\ and\
  \citenamefont {Doert}}]{baenitz_planar_2018}%
  \BibitemOpen
  \bibfield  {author} {\bibinfo {author} {\bibfnamefont {M.}~\bibnamefont
  {Baenitz}}, \bibinfo {author} {\bibfnamefont {P.}~\bibnamefont {Schlender}},
  \bibinfo {author} {\bibfnamefont {J.}~\bibnamefont {Sichelschmidt}}, \bibinfo
  {author} {\bibfnamefont {Y.~A.}\ \bibnamefont {Onykiienko}}, \bibinfo
  {author} {\bibfnamefont {Z.}~\bibnamefont {Zangeneh}}, \bibinfo {author}
  {\bibfnamefont {K.~M.}\ \bibnamefont {Ranjith}}, \bibinfo {author}
  {\bibfnamefont {R.}~\bibnamefont {Sarkar}}, \bibinfo {author} {\bibfnamefont
  {L.}~\bibnamefont {Hozoi}}, \bibinfo {author} {\bibfnamefont {H.~C.}\
  \bibnamefont {Walker}}, \bibinfo {author} {\bibfnamefont {J.-C.}\
  \bibnamefont {Orain}}, \bibinfo {author} {\bibfnamefont {H.}~\bibnamefont
  {Yasuoka}}, \bibinfo {author} {\bibfnamefont {J.}~\bibnamefont {van~den
  Brink}}, \bibinfo {author} {\bibfnamefont {H.~H.}\ \bibnamefont {Klauss}},
  \bibinfo {author} {\bibfnamefont {D.~S.}\ \bibnamefont {Inosov}}, \ and\
  \bibinfo {author} {\bibfnamefont {T.}~\bibnamefont {Doert}},\ }\href
  {\doibase 10.1103/PhysRevB.98.220409} {\bibfield  {journal} {\bibinfo
  {journal} {Phys. Rev. B}\ }\textbf {\bibinfo {volume} {98}},\ \bibinfo
  {pages} {220409(R)} (\bibinfo {year} {2018})}\BibitemShut {NoStop}%
\bibitem [{\citenamefont {Sichelschmidt}\ \emph {et~al.}(2019)\citenamefont
  {Sichelschmidt}, \citenamefont {Schlender}, \citenamefont {Schmidt},
  \citenamefont {Baenitz},\ and\ \citenamefont
  {Doert}}]{sichelschmidt_electron_2019}%
  \BibitemOpen
  \bibfield  {author} {\bibinfo {author} {\bibfnamefont {J.}~\bibnamefont
  {Sichelschmidt}}, \bibinfo {author} {\bibfnamefont {P.}~\bibnamefont
  {Schlender}}, \bibinfo {author} {\bibfnamefont {B.}~\bibnamefont {Schmidt}},
  \bibinfo {author} {\bibfnamefont {M.}~\bibnamefont {Baenitz}}, \ and\
  \bibinfo {author} {\bibfnamefont {T.}~\bibnamefont {Doert}},\ }\href
  {\doibase 10.1088/1361-648x/ab071d} {\bibfield  {journal} {\bibinfo
  {journal} {J. Phys.: Condens. Matter}\ }\textbf {\bibinfo {volume} {31}},\
  \bibinfo {pages} {205601} (\bibinfo {year} {2019})}\BibitemShut {NoStop}%
\bibitem [{\citenamefont {Bordelon}\ \emph {et~al.}(2019)\citenamefont
  {Bordelon}, \citenamefont {Kenney}, \citenamefont {Liu}, \citenamefont
  {Hogan}, \citenamefont {Posthuma}, \citenamefont {Kavand}, \citenamefont
  {Lyu}, \citenamefont {Sherwin}, \citenamefont {Butch}, \citenamefont {Brown},
  \citenamefont {Graf}, \citenamefont {Balents},\ and\ \citenamefont
  {Wilson}}]{bordelon_field_2019}%
  \BibitemOpen
  \bibfield  {author} {\bibinfo {author} {\bibfnamefont {M.~M.}\ \bibnamefont
  {Bordelon}}, \bibinfo {author} {\bibfnamefont {E.}~\bibnamefont {Kenney}},
  \bibinfo {author} {\bibfnamefont {C.}~\bibnamefont {Liu}}, \bibinfo {author}
  {\bibfnamefont {T.}~\bibnamefont {Hogan}}, \bibinfo {author} {\bibfnamefont
  {L.}~\bibnamefont {Posthuma}}, \bibinfo {author} {\bibfnamefont
  {M.}~\bibnamefont {Kavand}}, \bibinfo {author} {\bibfnamefont
  {Y.}~\bibnamefont {Lyu}}, \bibinfo {author} {\bibfnamefont {M.}~\bibnamefont
  {Sherwin}}, \bibinfo {author} {\bibfnamefont {N.~P.}\ \bibnamefont {Butch}},
  \bibinfo {author} {\bibfnamefont {C.}~\bibnamefont {Brown}}, \bibinfo
  {author} {\bibfnamefont {M.~J.}\ \bibnamefont {Graf}}, \bibinfo {author}
  {\bibfnamefont {L.}~\bibnamefont {Balents}}, \ and\ \bibinfo {author}
  {\bibfnamefont {S.~D.}\ \bibnamefont {Wilson}},\ }\href {\doibase
  10.1038/s41567-019-0594-5} {\bibfield  {journal} {\bibinfo  {journal} {Nat.
  Phys.}\ }\textbf {\bibinfo {volume} {15}},\ \bibinfo {pages} {1058} (\bibinfo
  {year} {2019})}\BibitemShut {NoStop}%
\bibitem [{\citenamefont {Ranjith}\ \emph {et~al.}(2019)\citenamefont
  {Ranjith}, \citenamefont {Dmytriieva}, \citenamefont {Khim}, \citenamefont
  {Sichelschmidt}, \citenamefont {Luther}, \citenamefont {Ehlers},
  \citenamefont {Yasuoka}, \citenamefont {Wosnitza}, \citenamefont {Tsirlin},
  \citenamefont {K\"uhne},\ and\ \citenamefont {Baenitz}}]{ranjith_field_2019}%
  \BibitemOpen
  \bibfield  {author} {\bibinfo {author} {\bibfnamefont {K.~M.}\ \bibnamefont
  {Ranjith}}, \bibinfo {author} {\bibfnamefont {D.}~\bibnamefont {Dmytriieva}},
  \bibinfo {author} {\bibfnamefont {S.}~\bibnamefont {Khim}}, \bibinfo {author}
  {\bibfnamefont {J.}~\bibnamefont {Sichelschmidt}}, \bibinfo {author}
  {\bibfnamefont {S.}~\bibnamefont {Luther}}, \bibinfo {author} {\bibfnamefont
  {D.}~\bibnamefont {Ehlers}}, \bibinfo {author} {\bibfnamefont
  {H.}~\bibnamefont {Yasuoka}}, \bibinfo {author} {\bibfnamefont
  {J.}~\bibnamefont {Wosnitza}}, \bibinfo {author} {\bibfnamefont {A.~A.}\
  \bibnamefont {Tsirlin}}, \bibinfo {author} {\bibfnamefont {H.}~\bibnamefont
  {K\"uhne}}, \ and\ \bibinfo {author} {\bibfnamefont {M.}~\bibnamefont
  {Baenitz}},\ }\href {\doibase 10.1103/PhysRevB.99.180401} {\bibfield
  {journal} {\bibinfo  {journal} {Phys. Rev. B}\ }\textbf {\bibinfo {volume}
  {99}},\ \bibinfo {pages} {180401(R)} (\bibinfo {year} {2019})}\BibitemShut
  {NoStop}%
\bibitem [{\citenamefont {Ding}\ \emph {et~al.}(2019)\citenamefont {Ding},
  \citenamefont {Manuel}, \citenamefont {Bachus}, \citenamefont {Gru\ss{}ler},
  \citenamefont {Gegenwart}, \citenamefont {Singleton}, \citenamefont
  {Johnson}, \citenamefont {Walker}, \citenamefont {Adroja}, \citenamefont
  {Hillier},\ and\ \citenamefont {Tsirlin}}]{ding_gapless_2019}%
  \BibitemOpen
  \bibfield  {author} {\bibinfo {author} {\bibfnamefont {L.}~\bibnamefont
  {Ding}}, \bibinfo {author} {\bibfnamefont {P.}~\bibnamefont {Manuel}},
  \bibinfo {author} {\bibfnamefont {S.}~\bibnamefont {Bachus}}, \bibinfo
  {author} {\bibfnamefont {F.}~\bibnamefont {Gru\ss{}ler}}, \bibinfo {author}
  {\bibfnamefont {P.}~\bibnamefont {Gegenwart}}, \bibinfo {author}
  {\bibfnamefont {J.}~\bibnamefont {Singleton}}, \bibinfo {author}
  {\bibfnamefont {R.~D.}\ \bibnamefont {Johnson}}, \bibinfo {author}
  {\bibfnamefont {H.~C.}\ \bibnamefont {Walker}}, \bibinfo {author}
  {\bibfnamefont {D.~T.}\ \bibnamefont {Adroja}}, \bibinfo {author}
  {\bibfnamefont {A.~D.}\ \bibnamefont {Hillier}}, \ and\ \bibinfo {author}
  {\bibfnamefont {A.~A.}\ \bibnamefont {Tsirlin}},\ }\href {\doibase
  10.1103/PhysRevB.100.144432} {\bibfield  {journal} {\bibinfo  {journal}
  {Phys. Rev. B}\ }\textbf {\bibinfo {volume} {100}},\ \bibinfo {pages}
  {144432} (\bibinfo {year} {2019})}\BibitemShut {NoStop}%
\bibitem [{\citenamefont {Gardner}\ \emph {et~al.}(2010)\citenamefont
  {Gardner}, \citenamefont {Gingras},\ and\ \citenamefont
  {Greedan}}]{gardner_magnetic_2010}%
  \BibitemOpen
  \bibfield  {author} {\bibinfo {author} {\bibfnamefont {J.~S.}\ \bibnamefont
  {Gardner}}, \bibinfo {author} {\bibfnamefont {M.~J.~P.}\ \bibnamefont
  {Gingras}}, \ and\ \bibinfo {author} {\bibfnamefont {J.~E.}\ \bibnamefont
  {Greedan}},\ }\href {\doibase 10.1103/RevModPhys.82.53} {\bibfield  {journal}
  {\bibinfo  {journal} {Rev. Mod. Phys.}\ }\textbf {\bibinfo {volume} {82}},\
  \bibinfo {pages} {53} (\bibinfo {year} {2010})}\BibitemShut {NoStop}%
\bibitem [{\citenamefont {{A Bertin and Y Chapuis and P Dalmas de Réotier and
  A Yaouanc}}(2012)}]{bertin_crystal_2012}%
  \BibitemOpen
  \bibfield  {author} {\bibinfo {author} {\bibnamefont {{A Bertin and Y Chapuis
  and P Dalmas de Réotier and A Yaouanc}}},\ }\href {\doibase
  10.1088/0953-8984/24/25/256003} {\bibfield  {journal} {\bibinfo  {journal}
  {J. Phys.: Condens. Matter}\ }\textbf {\bibinfo {volume} {24}},\ \bibinfo
  {pages} {256003} (\bibinfo {year} {2012})}\BibitemShut {NoStop}%
\bibitem [{\citenamefont {Gao}\ \emph {et~al.}(2018)\citenamefont {Gao},
  \citenamefont {Zaharko}, \citenamefont {Tsurkan}, \citenamefont {Prodan},
  \citenamefont {Riordan}, \citenamefont {Lago}, \citenamefont {F{\aa}k},
  \citenamefont {Wildes}, \citenamefont {Koza}, \citenamefont {Ritter},
  \citenamefont {Fouquet}, \citenamefont {Keller}, \citenamefont {Can\'evet},
  \citenamefont {Medarde}, \citenamefont {Blomgren}, \citenamefont {Johansson},
  \citenamefont {Giblin}, \citenamefont {Vrtnik}, \citenamefont {Luzar},
  \citenamefont {Loidl}, \citenamefont {R\"uegg},\ and\ \citenamefont
  {Fennell}}]{gao_dipolar_2018}%
  \BibitemOpen
  \bibfield  {author} {\bibinfo {author} {\bibfnamefont {S.}~\bibnamefont
  {Gao}}, \bibinfo {author} {\bibfnamefont {O.}~\bibnamefont {Zaharko}},
  \bibinfo {author} {\bibfnamefont {V.}~\bibnamefont {Tsurkan}}, \bibinfo
  {author} {\bibfnamefont {L.}~\bibnamefont {Prodan}}, \bibinfo {author}
  {\bibfnamefont {E.}~\bibnamefont {Riordan}}, \bibinfo {author} {\bibfnamefont
  {J.}~\bibnamefont {Lago}}, \bibinfo {author} {\bibfnamefont {B.}~\bibnamefont
  {F{\aa}k}}, \bibinfo {author} {\bibfnamefont {A.~R.}\ \bibnamefont {Wildes}},
  \bibinfo {author} {\bibfnamefont {M.~M.}\ \bibnamefont {Koza}}, \bibinfo
  {author} {\bibfnamefont {C.}~\bibnamefont {Ritter}}, \bibinfo {author}
  {\bibfnamefont {P.}~\bibnamefont {Fouquet}}, \bibinfo {author} {\bibfnamefont
  {L.}~\bibnamefont {Keller}}, \bibinfo {author} {\bibfnamefont
  {E.}~\bibnamefont {Can\'evet}}, \bibinfo {author} {\bibfnamefont
  {M.}~\bibnamefont {Medarde}}, \bibinfo {author} {\bibfnamefont
  {J.}~\bibnamefont {Blomgren}}, \bibinfo {author} {\bibfnamefont
  {C.}~\bibnamefont {Johansson}}, \bibinfo {author} {\bibfnamefont {S.~R.}\
  \bibnamefont {Giblin}}, \bibinfo {author} {\bibfnamefont {S.}~\bibnamefont
  {Vrtnik}}, \bibinfo {author} {\bibfnamefont {J.}~\bibnamefont {Luzar}},
  \bibinfo {author} {\bibfnamefont {A.}~\bibnamefont {Loidl}}, \bibinfo
  {author} {\bibfnamefont {C.}~\bibnamefont {R\"uegg}}, \ and\ \bibinfo
  {author} {\bibfnamefont {T.}~\bibnamefont {Fennell}},\ }\href {\doibase
  10.1103/PhysRevLett.120.137201} {\bibfield  {journal} {\bibinfo  {journal}
  {Phys. Rev. Lett.}\ }\textbf {\bibinfo {volume} {120}},\ \bibinfo {pages}
  {137201} (\bibinfo {year} {2018})}\BibitemShut {NoStop}%
\bibitem [{\citenamefont {Reig-i Plessis}\ \emph {et~al.}(2019)\citenamefont
  {Reig-i Plessis}, \citenamefont {Cote}, \citenamefont {van Geldern},
  \citenamefont {Mayrhofer}, \citenamefont {Aczel},\ and\ \citenamefont
  {MacDougall}}]{reig_neutron_2019}%
  \BibitemOpen
  \bibfield  {author} {\bibinfo {author} {\bibfnamefont {D.}~\bibnamefont
  {Reig-i Plessis}}, \bibinfo {author} {\bibfnamefont {A.}~\bibnamefont
  {Cote}}, \bibinfo {author} {\bibfnamefont {S.}~\bibnamefont {van Geldern}},
  \bibinfo {author} {\bibfnamefont {R.~D.}\ \bibnamefont {Mayrhofer}}, \bibinfo
  {author} {\bibfnamefont {A.~A.}\ \bibnamefont {Aczel}}, \ and\ \bibinfo
  {author} {\bibfnamefont {G.~J.}\ \bibnamefont {MacDougall}},\ }\href@noop {}
  {\bibfield  {journal} {\bibinfo  {journal} {arXiv:1906.10767}\ } (\bibinfo
  {year} {2019})}\BibitemShut {NoStop}%
\bibitem [{\citenamefont {Rodriguez-Carvajal}(1993)}]{rodriguez_recent_1993}%
  \BibitemOpen
  \bibfield  {author} {\bibinfo {author} {\bibfnamefont {J.}~\bibnamefont
  {Rodriguez-Carvajal}},\ }\href@noop {} {\bibfield  {journal} {\bibinfo
  {journal} {Physica B: Conden. Matter}\ }\textbf {\bibinfo {volume} {192}},\
  \bibinfo {pages} {55} (\bibinfo {year} {1993})}\BibitemShut {NoStop}%
\bibitem [{\citenamefont {Kajimoto}\ \emph {et~al.}(2011)\citenamefont
  {Kajimoto}, \citenamefont {Nakamura}, \citenamefont {Inamura}, \citenamefont
  {Mizuno}, \citenamefont {Nakajima}, \citenamefont {Ohira-Kawamura},
  \citenamefont {Yokoo}, \citenamefont {Nakatani}, \citenamefont {Maruyama},
  \citenamefont {Soyama}, \citenamefont {Shibata}, \citenamefont {Suzuya},
  \citenamefont {Sato}, \citenamefont {Aizawa}, \citenamefont {Arai},
  \citenamefont {Wakimoto}, \citenamefont {Ishikado}, \citenamefont {Shamoto},
  \citenamefont {Fujita}, \citenamefont {Hiraka}, \citenamefont {Ohoyama},
  \citenamefont {Yamada},\ and\ \citenamefont {Lee}}]{kajimoto_fermi_2011}%
  \BibitemOpen
  \bibfield  {author} {\bibinfo {author} {\bibfnamefont {R.}~\bibnamefont
  {Kajimoto}}, \bibinfo {author} {\bibfnamefont {M.}~\bibnamefont {Nakamura}},
  \bibinfo {author} {\bibfnamefont {Y.}~\bibnamefont {Inamura}}, \bibinfo
  {author} {\bibfnamefont {F.}~\bibnamefont {Mizuno}}, \bibinfo {author}
  {\bibfnamefont {K.}~\bibnamefont {Nakajima}}, \bibinfo {author}
  {\bibfnamefont {S.}~\bibnamefont {Ohira-Kawamura}}, \bibinfo {author}
  {\bibfnamefont {T.}~\bibnamefont {Yokoo}}, \bibinfo {author} {\bibfnamefont
  {T.}~\bibnamefont {Nakatani}}, \bibinfo {author} {\bibfnamefont
  {R.}~\bibnamefont {Maruyama}}, \bibinfo {author} {\bibfnamefont
  {K.}~\bibnamefont {Soyama}}, \bibinfo {author} {\bibfnamefont
  {K.}~\bibnamefont {Shibata}}, \bibinfo {author} {\bibfnamefont
  {K.}~\bibnamefont {Suzuya}}, \bibinfo {author} {\bibfnamefont
  {S.}~\bibnamefont {Sato}}, \bibinfo {author} {\bibfnamefont {K.}~\bibnamefont
  {Aizawa}}, \bibinfo {author} {\bibfnamefont {M.}~\bibnamefont {Arai}},
  \bibinfo {author} {\bibfnamefont {S.}~\bibnamefont {Wakimoto}}, \bibinfo
  {author} {\bibfnamefont {M.}~\bibnamefont {Ishikado}}, \bibinfo {author}
  {\bibfnamefont {S.-i.}\ \bibnamefont {Shamoto}}, \bibinfo {author}
  {\bibfnamefont {M.}~\bibnamefont {Fujita}}, \bibinfo {author} {\bibfnamefont
  {H.}~\bibnamefont {Hiraka}}, \bibinfo {author} {\bibfnamefont
  {K.}~\bibnamefont {Ohoyama}}, \bibinfo {author} {\bibfnamefont
  {K.}~\bibnamefont {Yamada}}, \ and\ \bibinfo {author} {\bibfnamefont {C.-H.}\
  \bibnamefont {Lee}},\ }\href {\doibase 10.1143/JPSJS.80SB.SB025} {\bibfield
  {journal} {\bibinfo  {journal} {J. Phys. Soc. Jpn.}\ }\textbf {\bibinfo
  {volume} {80}},\ \bibinfo {pages} {SB025} (\bibinfo {year}
  {2011})}\BibitemShut {NoStop}%
\bibitem [{\citenamefont {Ruminy}\ \emph {et~al.}(2016)\citenamefont {Ruminy},
  \citenamefont {Pomjakushina}, \citenamefont {Iida}, \citenamefont {Kamazawa},
  \citenamefont {Adroja}, \citenamefont {Stuhr},\ and\ \citenamefont
  {Fennell}}]{ruminy_crystal_2016}%
  \BibitemOpen
  \bibfield  {author} {\bibinfo {author} {\bibfnamefont {M.}~\bibnamefont
  {Ruminy}}, \bibinfo {author} {\bibfnamefont {E.}~\bibnamefont
  {Pomjakushina}}, \bibinfo {author} {\bibfnamefont {K.}~\bibnamefont {Iida}},
  \bibinfo {author} {\bibfnamefont {K.}~\bibnamefont {Kamazawa}}, \bibinfo
  {author} {\bibfnamefont {D.~T.}\ \bibnamefont {Adroja}}, \bibinfo {author}
  {\bibfnamefont {U.}~\bibnamefont {Stuhr}}, \ and\ \bibinfo {author}
  {\bibfnamefont {T.}~\bibnamefont {Fennell}},\ }\href {\doibase
  10.1103/PhysRevB.94.024430} {\bibfield  {journal} {\bibinfo  {journal} {Phys.
  Rev. B}\ }\textbf {\bibinfo {volume} {94}},\ \bibinfo {pages} {024430}
  (\bibinfo {year} {2016})}\BibitemShut {NoStop}%
\bibitem [{\citenamefont {Nakamura}\ \emph {et~al.}(2009)\citenamefont
  {Nakamura}, \citenamefont {Kajimoto}, \citenamefont {Inamura}, \citenamefont
  {Mizuno}, \citenamefont {Fujita}, \citenamefont {Yokoo},\ and\ \citenamefont
  {Arai}}]{nakamura_first_2009}%
  \BibitemOpen
  \bibfield  {author} {\bibinfo {author} {\bibfnamefont {M.}~\bibnamefont
  {Nakamura}}, \bibinfo {author} {\bibfnamefont {R.}~\bibnamefont {Kajimoto}},
  \bibinfo {author} {\bibfnamefont {Y.}~\bibnamefont {Inamura}}, \bibinfo
  {author} {\bibfnamefont {F.}~\bibnamefont {Mizuno}}, \bibinfo {author}
  {\bibfnamefont {M.}~\bibnamefont {Fujita}}, \bibinfo {author} {\bibfnamefont
  {T.}~\bibnamefont {Yokoo}}, \ and\ \bibinfo {author} {\bibfnamefont
  {M.}~\bibnamefont {Arai}},\ }\href {\doibase 10.1143/JPSJ.78.093002}
  {\bibfield  {journal} {\bibinfo  {journal} {J. Phys. Soc. Jpn.}\ }\textbf
  {\bibinfo {volume} {78}},\ \bibinfo {pages} {093002} (\bibinfo {year}
  {2009})}\BibitemShut {NoStop}%
\bibitem [{\citenamefont {Inamura}\ \emph {et~al.}(2013)\citenamefont
  {Inamura}, \citenamefont {Nakatani}, \citenamefont {Suzuki},\ and\
  \citenamefont {Otomo}}]{inamura_development_2013}%
  \BibitemOpen
  \bibfield  {author} {\bibinfo {author} {\bibfnamefont {Y.}~\bibnamefont
  {Inamura}}, \bibinfo {author} {\bibfnamefont {T.}~\bibnamefont {Nakatani}},
  \bibinfo {author} {\bibfnamefont {J.}~\bibnamefont {Suzuki}}, \ and\ \bibinfo
  {author} {\bibfnamefont {T.}~\bibnamefont {Otomo}},\ }\href {\doibase
  10.7566/JPSJS.82SA.SA031} {\bibfield  {journal} {\bibinfo  {journal} {J.
  Phys. Soc. Jpn.}\ }\textbf {\bibinfo {volume} {82}},\ \bibinfo {pages}
  {SA031} (\bibinfo {year} {2013})}\BibitemShut {NoStop}%
\bibitem [{\citenamefont {Hutchings}(1964)}]{hutchings_point_1964}%
  \BibitemOpen
  \bibfield  {author} {\bibinfo {author} {\bibfnamefont {M.}~\bibnamefont
  {Hutchings}}\ }(\bibinfo  {publisher} {Academic Press},\ \bibinfo {year}
  {1964})\ pp.\ \bibinfo {pages} {227 -- 273}\BibitemShut {NoStop}%
\bibitem [{\citenamefont {Rotter}(2004)}]{rotter_using_2004}%
  \BibitemOpen
  \bibfield  {author} {\bibinfo {author} {\bibfnamefont {M.}~\bibnamefont
  {Rotter}},\ }\href {\doibase 10.1016/j.jmmm.2003.12.1394} {\bibfield
  {journal} {\bibinfo  {journal} {J. Mag. Magn. Mater.}\ }\textbf {\bibinfo
  {volume} {272–276}},\ \bibinfo {pages} {E481} (\bibinfo {year}
  {2004})}\BibitemShut {NoStop}%
\bibitem [{\citenamefont {Rosenkranz}\ \emph {et~al.}(2000)\citenamefont
  {Rosenkranz}, \citenamefont {Ramirez}, \citenamefont {Hayashi}, \citenamefont
  {Cava}, \citenamefont {Siddharthan},\ and\ \citenamefont
  {Shastry}}]{rosenkranz_crystal_2000}%
  \BibitemOpen
  \bibfield  {author} {\bibinfo {author} {\bibfnamefont {S.}~\bibnamefont
  {Rosenkranz}}, \bibinfo {author} {\bibfnamefont {A.~P.}\ \bibnamefont
  {Ramirez}}, \bibinfo {author} {\bibfnamefont {A.}~\bibnamefont {Hayashi}},
  \bibinfo {author} {\bibfnamefont {R.~J.}\ \bibnamefont {Cava}}, \bibinfo
  {author} {\bibfnamefont {R.}~\bibnamefont {Siddharthan}}, \ and\ \bibinfo
  {author} {\bibfnamefont {B.~S.}\ \bibnamefont {Shastry}},\ }\href {\doibase
  10.1063/1.372565} {\bibfield  {journal} {\bibinfo  {journal} {J. Appl.
  Phys.}\ }\textbf {\bibinfo {volume} {87}},\ \bibinfo {pages} {5914} (\bibinfo
  {year} {2000})}\BibitemShut {NoStop}%
\bibitem [{\citenamefont {Xu}\ \emph {et~al.}(2015)\citenamefont {Xu},
  \citenamefont {Anand}, \citenamefont {Bera}, \citenamefont {Frontzek},
  \citenamefont {Abernathy}, \citenamefont {Casati}, \citenamefont
  {Siemensmeyer},\ and\ \citenamefont {Lake}}]{xu_magnetic_2015}%
  \BibitemOpen
  \bibfield  {author} {\bibinfo {author} {\bibfnamefont {J.}~\bibnamefont
  {Xu}}, \bibinfo {author} {\bibfnamefont {V.~K.}\ \bibnamefont {Anand}},
  \bibinfo {author} {\bibfnamefont {A.~K.}\ \bibnamefont {Bera}}, \bibinfo
  {author} {\bibfnamefont {M.}~\bibnamefont {Frontzek}}, \bibinfo {author}
  {\bibfnamefont {D.~L.}\ \bibnamefont {Abernathy}}, \bibinfo {author}
  {\bibfnamefont {N.}~\bibnamefont {Casati}}, \bibinfo {author} {\bibfnamefont
  {K.}~\bibnamefont {Siemensmeyer}}, \ and\ \bibinfo {author} {\bibfnamefont
  {B.}~\bibnamefont {Lake}},\ }\href {\doibase 10.1103/PhysRevB.92.224430}
  {\bibfield  {journal} {\bibinfo  {journal} {Phys. Rev. B}\ }\textbf {\bibinfo
  {volume} {92}},\ \bibinfo {pages} {224430} (\bibinfo {year}
  {2015})}\BibitemShut {NoStop}%
\bibitem [{\citenamefont {Le}()}]{le_saficf}%
  \BibitemOpen
  \bibfield  {author} {\bibinfo {author} {\bibfnamefont {M.~D.}\ \bibnamefont
  {Le}},\ }\enquote {\bibinfo {title} {http://yocto.me/work/saficf/index.html}}\ \BibitemShut {NoStop}%
\bibitem [{\citenamefont {Li}\ \emph {et~al.}(2017)\citenamefont {Li},
  \citenamefont {Adroja}, \citenamefont {Bewley}, \citenamefont {Voneshen},
  \citenamefont {Tsirlin}, \citenamefont {Gegenwart},\ and\ \citenamefont
  {Zhang}}]{li_crystalline_2017}%
  \BibitemOpen
  \bibfield  {author} {\bibinfo {author} {\bibfnamefont {Y.}~\bibnamefont
  {Li}}, \bibinfo {author} {\bibfnamefont {D.}~\bibnamefont {Adroja}}, \bibinfo
  {author} {\bibfnamefont {R.~I.}\ \bibnamefont {Bewley}}, \bibinfo {author}
  {\bibfnamefont {D.}~\bibnamefont {Voneshen}}, \bibinfo {author}
  {\bibfnamefont {A.~A.}\ \bibnamefont {Tsirlin}}, \bibinfo {author}
  {\bibfnamefont {P.}~\bibnamefont {Gegenwart}}, \ and\ \bibinfo {author}
  {\bibfnamefont {Q.}~\bibnamefont {Zhang}},\ }\href {\doibase
  10.1103/PhysRevLett.118.107202} {\bibfield  {journal} {\bibinfo  {journal}
  {Phys. Rev. Lett.}\ }\textbf {\bibinfo {volume} {118}},\ \bibinfo {pages}
  {107202} (\bibinfo {year} {2017})}\BibitemShut {NoStop}%
\bibitem [{\citenamefont {Scheie}\ \emph {et~al.}(2020)\citenamefont {Scheie},
  \citenamefont {Garlea}, \citenamefont {Sanjeewa}, \citenamefont {Xing},\ and\
  \citenamefont {Sefat}}]{scheie_crystal_2020}%
  \BibitemOpen
  \bibfield  {author} {\bibinfo {author} {\bibfnamefont {A.}~\bibnamefont
  {Scheie}}, \bibinfo {author} {\bibfnamefont {V.~O.}\ \bibnamefont {Garlea}},
  \bibinfo {author} {\bibfnamefont {L.~D.}\ \bibnamefont {Sanjeewa}}, \bibinfo
  {author} {\bibfnamefont {J.}~\bibnamefont {Xing}}, \ and\ \bibinfo {author}
  {\bibfnamefont {A.~S.}\ \bibnamefont {Sefat}},\ }\href {\doibase
  10.1103/PhysRevB.101.144432} {\bibfield  {journal} {\bibinfo  {journal}
  {Phys. Rev. B}\ }\textbf {\bibinfo {volume} {101}},\ \bibinfo {pages}
  {144432} (\bibinfo {year} {2020})}\BibitemShut {NoStop}%
\end{thebibliography}
\end{document}